\documentclass[12pt]{article}
\usepackage[margin=1in]{geometry}
\usepackage[utf8]{inputenc}

\usepackage[font=small,labelfont=bf]{caption}

\usepackage{graphicx} 
\usepackage{multirow, multicol, makecell}

\usepackage{amsmath,amssymb,amsfonts,amsthm}

\usepackage{xcolor}
\usepackage{url}
\usepackage{hyperref}
\hypersetup{
  colorlinks   = true, 
  urlcolor     = blue, 
  linkcolor    = black, 
  citecolor   = black 
}
\usepackage[numbers,sort&compress]{natbib}
\let\OLDthebibliography\thebibliography
\renewcommand\thebibliography[1]{
  \OLDthebibliography{#1}
  \setlength{\parskip}{4pt}
  \setlength{\itemsep}{0pt plus 0.3ex}
}

\allowdisplaybreaks
\usepackage{tabularx}
\usepackage{bbm}
\usepackage{listings}
\usepackage{soul}
\usepackage{feynmp-auto}
\usepackage{bbold}

\newcommand\refeq[1]{Eq.~(\ref{#1})}

\newcommand\refta[1]{Tab.~\ref{#1}}
\newcommand\refse[1]{Sect.~\ref{#1}}

\newcommand\citere[1]{Ref.~\cite{#1}}
\newcommand\citeres[1]{Refs.~\cite{#1}}
\newcommand\refap[1]{App.~\ref{#1}}
\def\reffi#1{\mbox{Fig.~\ref{#1}}}

\newcommand{\gev}{\ \mathrm{GeV}}
\newcommand{\tev}{\ \mathrm{TeV}}
\newcommand{\tevinv}{\ \mathrm{TeV}^{-1}}

\newcommand{\fb}{\mathrm{fb}^{-1}}
\newcommand{\ab}{\mathrm{ab}^{-1}}
\newcommand{\ttbar}{{t \bar{t}}}
\newcommand{\mtt}{m_{\ttbar}}
\newcommand{\chel}{c_{\mathrm{hel}}}
\newcommand{\cG}{c_{{\tilde{G}}}}
\newcommand{\ct}{c_{t}}
\newcommand{\gAtt}{g_{{A \ttbar}}}
\newcommand{\ma}{m_a}
\newcommand{\ga}{\Gamma_a}

\newcommand{\flavor}{flavor}
\newcommand{\program}{program}
\newcommand{\colour}{color}
\newcommand{\analyze}{analyze}

\begin{document}

\thispagestyle{empty}

\def\thefootnote{\fnsymbol{footnote}}
\begin{flushright}
  \footnotesize
  DESY-24-059 \qquad IFT--UAM/CSIC-24-042 \qquad KA-TP-06-2024 \qquad MITP-24-044 
\end{flushright}

\begin{center}

{\Large\textbf{
ALP-ine quests at the LHC: hunting axion-like\\[0.2em] particles via peaks
and dips  in
\boldmath{$\ttbar$} production
}}

\vspace*{2em}

Afiq Anuar$^1$, 
Anke Biek\"otter$^2$, 
Thomas Biek\"otter$^{3,5}$, 
Alexander Grohsjean$^4$, 
Sven~Heinemeyer$^5$, 
Laurids Jeppe$^1$, 
Christian Schwanenberger$^{1,4}$, 
and
Georg Weiglein$^{1,6}$\footnote{Emails: afiq.anuar@cern.ch, biekoetter@uni-mainz.de, thomas.biekoetter@desy.de,\\ alexander.grohsjean@desy.de, sven.heinemeyer@cern.ch, laurids.jeppe@desy.de,\\ christian.schwanenberger@desy.de, georg.weiglein@desy.de}

\vspace*{0.4em}

\textit{
$^1$Deutsches Elektronen-Synchrotron DESY,
Notkestr.~85, 22607 Hamburg, Germany\\[0.2em]
$^2$PRISMA+ Cluster of Excellence \& Institute of Physics~(THEP) \& Mainz Institute for Theoretical
Physics, Johannes Gutenberg University, D-55099 Mainz, Germany\\[0.2em]
$^3$Institute for Theoretical Physics,
Karlsruhe Institute of Technology,\\
Wolfgang-Gaede-Str.~1, 76131 Karlsruhe, Germany\\[0.2em]
$^4$ Institut f\"ur Experimentalphysik, Universit\"at Hamburg,\\
Luruper Chaussee 149, 22761 Hamburg, Germany\\[0.2em]
$^5$Instituto de Física Teórica (UAM/CSIC), Cantoblanco, 28049,
Madrid, Spain\\[0.4em]
$^6$II.~Institut f\"ur Theoretische Physik, Universit\"at Hamburg,\\
Luruper Chaussee 149, 22761 Hamburg, Germany
}

\vspace*{0.2cm}

\begin{abstract}
We present an analysis of the sensitivity
of current and future LHC searches for new spin-0 particles
in
top--anti-top-quark ($t\bar{t}$) final states, focusing on
generic axion-like particles~(ALPs) that are
coupled to top quarks and gluons.
As a first step, we derive new limits on the effective ALP
Lagrangian in terms of the Wilson
coefficients $\ct$ and $\cG$
based on the 
results of the
CMS search using $35.9$~fb$^{-1}$ of data, collected
at $\sqrt{s} = 13$~TeV.
We then investigate how the
production of an ALP
with generic couplings to gluons and top quarks 
can be distinguished 
from the production of a pseudoscalar
which couples to gluons exclusively via a top-quark loop.
To this end, we make use of the invariant $t\bar{t}$ 
mass distribution and angular correlations that are sensitive to the $t\bar{t}$ spin 
correlation.
Using a mass of~400~GeV as an example, 
we find that 
already the data collected during Run~2 and Run~3 of the LHC provides
an interesting sensitivity to the underlying nature of a possible 
new particle.
We also analyze the prospects for data
anticipated to be collected during the
high-luminosity phase
of the~LHC.
Finally, we compare 
the limits 
obtained from the $t \bar t$ searches
to existing experimental
bounds from 
LHC searches
for narrow di-photon resonances,
from 
measurements
of the production of four top quarks,
and from global analyses of ALP--SMEFT
interference effects.

\end{abstract}

\end{center}
\renewcommand{\thefootnote}{\arabic{footnote}}
\setcounter{footnote}{0}

\newpage

\tableofcontents

\section{Introduction}

Axions and axion-like particles (ALPs, denoted $a$ in the following) are spin-0 particles that are singlets under the Standard Model (SM) gauge groups.
ALPs appear in many well-motivated SM extensions, 
where they arise as pseudo-Nambu-Goldstone bosons of
an approximate axion shift-symmetry.
As a consequence, the masses of ALPs can naturally be much smaller than the energy scale of
the underlying ultraviolet (UV) model, making them an attractive target for the Large 
Hadron Collider (LHC) and the future High-Luminosity LHC (HL-LHC).
While axions have originally been
introduced as a potential solution to
the strong-CP problem~\cite{Peccei:1977hh,
Peccei:1977ur,
Weinberg:1977ma,Wilczek:1977pj}, ALPs are
featured in a variety of
SM extensions~\cite{DiLuzio:2020wdo,Choi:2020rgn,
Chadha-Day:2021szb}
including string
theory~\cite{Witten:1984dg,Svrcek:2006yi},
supersymmetric
theories~\cite{Frere:1983ag},
dark-matter models~\cite{Preskill:1982cy,
Abbott:1982af,Dine:1982ah}
and composite Higgs
models~\cite{Katz:2005au}.

Early analyses have focused on generic ALPs with masses below the GeV-range. 
However, also heavier
ALPs with masses of tens or hundreds
of~GeV that can be resonantly produced
at colliders are under active investigation, both in the $pp$~\cite{ATLAS:2020ahi,ATLAS:2021hbr,CMS:2018nsh,ATLAS:2020pcy} and $\gamma \gamma$ (light-by-light scattering)~\cite{ATLAS:2020hii,CMS:2018erd} production channels.
In particular, the so-called QCD axion addressing
the strong-CP problem can have a mass
in the TeV range if its mass receives additional
contributions from the confinement scale
associated with extra non-abelian gauge
groups~\cite{Rubakov:1997vp,Holdom:1982ex},
making it potentially accessible at the
LHC~\cite{Dimopoulos:2016lvn,Gherghetta:2016fhp}.
In this context, one should note that
such \textit{heavy QCD axions} are less prone
to the so-called axion quality
problem~\cite{Georgi:1981pu,
Lazarides:1985bj,
Holman:1992us,Barr:1992qq,
Ghigna:1992iv,
Kamionkowski:1992mf}
of the usual Peccei-Quinn mechanism,
such that they are also denoted
\textit{high-quality} axions
in recent literature~\cite{Hook:2019qoh}.\footnote{High-mass
axions can be exposed to other forms
of \textit{heavy axion quality
problems}~\cite{Valenti:2022tsc},
e.g.~associated with external sources of
CP violation~\cite{Bedi:2022qrd,Bedi:2024wqg},
see also
\citere{Bonnefoy:2022vop}.}
This further motivates searches for ALPs at the
LHC and future accelerators.

In recent studies, limits on ALP couplings
arising from existing collider searches
have been investigated with a main focus
on the ALP couplings
to the SM gauge bosons, both
for resonant~\cite{Mimasu:2014nea,
Jaeckel:2015jla,Brivio:2017ije,
Bauer:2017ris,Craig:2018kne,Bauer:2021mvw}
and non-resonant ALP
contributions~\cite{Gavela:2019cmq,
Carra:2021ycg,Galda:2021hbr,Biekotter:2023mpd,Biswas:2023ksj}.
However, generic ALPs are also expected
to be coupled to the SM fermions at the
electroweak~(EW) scale, e.g.\ via 
contributions that are induced by renormalisation-group running even if ALP--fermion
couplings are absent in the UV~\cite{Choi:2017gpf,MartinCamalich:2020dfe,Chala:2020wvs,Bauer:2020jbp}.
ALP--fermion couplings are typically assumed to have a \flavor-hierarchical structure~\cite{Froggatt:1978nt,Georgi:1986df,Gherghetta:2000qt,Agashe:2004rs}.
Moreover, the couplings of ALPs to fermions are typically proportional
to the fermion masses.
This results in a particular relevance of the ALP coupling to top quarks. 
Interactions between the ALP and top quarks
are also motivated based on naturalness
arguments and (non-minimal) composite Higgs
models~\cite{Katz:2005au,Gripaios:2009pe}.
Limits on the ALP--top-quark coupling have been derived from $t \Bar{t} a$
searches, from the effects of ALPs on $\ttbar$, $\ttbar \ttbar$ and $\ttbar b\bar b$ and $tja$ production~\cite{Esser:2023fdo, Blasi:2023hvb,Bruggisser:2023npd, Phan:2023dqw,Cheung:2024wve} as well as from
renormalization group~(RG) running effects on observables beyond those involving top quarks at the LHC~\cite{Bauer:2017ris}. 
Constraints from low-energy colliders for ALPs 
have been derived in Ref.~\cite{Bauer:2021mvw}.
Moreover, new searches for ALPs at
existing~\cite{Ferber:2022rsf,Behr:2023nch,Rygaard:2023vyo}
and future colliders~\cite{Hosseini:2022tac,
Chigusa:2023rrz,Yue:2023mjm}
have been proposed.

In this work (see~\citere{Biekotter:2024mpv} for preliminary results), we address ALP contributions to $\ttbar$ production in the dilepton decay channel for ALPs with masses above the $\ttbar$ threshold. 
The possibility to search for new $s$-channel resonances
in the $t \bar t$ invariant mass distribution at the
LHC has been studied in \citeres{Gaemers:1984sj,
Dicus:1994bm,Bernreuther:1997gs,Frederix:2007gi,Carena:2016npr,Djouadi:2019cbm,Bahl:2024fjb},
emphasizing the importance of signal--background
interference effects on the shape of the $m_{t \bar t}$
distribution and the resulting
characteristic ``peak-dip'' structures.
A first search taking into account the interference with the SM background for scalar $\ttbar$ production  has been published by the ATLAS collaboration using $20.3~\fb$ of 8~TeV $pp$ collisions~\cite{ATLAS:2017snw}. Exploring $35.9~\fb$ of 13~TeV LHC data, the CMS collaboration could further enhance the sensitivity to $\ttbar$ production via scalar and pseudoscalar resonances~\cite{CMS:2019pzc}.
During the final stages of preparation of this work, ATLAS published a result based on the full Run~2 data set, yielding slightly stronger expected constraints on the coupling between top quarks and the scalar/pseudoscalar boson~\cite{ATLAS:2024vxm}.

We perform a reinterpretation of the published CMS search for pseudoscalars in terms of ALPs to $t\bar{t}$ production
and extend it
by considering an ALP with a more general coupling structure which features an additional (besides the contribution induced by the top-quark loop)
effective coupling to gluons. 
Such an additional contribution to the effective
gluon coupling could originate, for instance, from
heavy vector-like quarks or from colored scalars
predicted in Supersymmetry, as studied
in \citeres{Carena:2016npr,Djouadi:2019cbm}.
We address the question how an ALP with both top-quark and gluon couplings could be distinguished from the case where the coupling of the new particle to gluons
is induced exclusively through the SM-like top-quark loop. 
We will refer to the second, more restrictive scenario as a \textit{pseudoscalar Higgs boson},
denoted as $A$, as it could result from models extending the SM only in the Higgs sector,
such as the Two Higgs doublet model
(2HDM)~\cite{Lee:1973iz,Kim:1979if,Wilczek:1977pj,
Branco:2011iw}.\footnote{If additional Higgs
bosons contained in the 2HDM
(or any other UV-complete theory)
can be produced at the LHC, one would 
of course have a
further possibility to distinguish between
such a model and the ALP framework in which other
BSM states are assumed to be much heavier.
In this paper, we do not consider
the impact of the additional Higgs bosons
present in the 2HDM since we study the
question to what extent a distinction
is possible betweeen the production of the
2HDM pseudoscalar $A$ and an ALP $a$
based only 
on the presence of a signal in
the $A/a \to t \bar t$ searches.}
Note that we only use the terms ``ALP" and ``pseudoscalar Higgs boson" to distinguish between the scenarios with more general and restrictive couplings structures, respectively. In principle, ALPs and pseudoscalar Higgs bosons could feature both 
coupling structures.\footnote{Even
in the 2HDM the pseudoscalar Higgs boson
obtains additional contributions to
the gluon coupling from the lighter quarks
which can become significant for large
values of $\tan\beta$. However, at large
$\tan\beta$ the coupling to top quarks
is suppressed. As a result,
if these additional contributions
are relevant,
the LHC searches in
the $t \bar t$ final state have no
sensitivity to the presence of the
additional Higgs bosons of the 2HDM.}

Our analysis is based on the invariant $t\bar{t}$ mass distribution. 
As our study involves the leptons from the top-quark decay, thereby going beyond the analysis level of stable top  quarks as considered, e.g., in \citere{Carena:2016npr,Phan:2023dqw}, 
we are able to employ angular variables. 
Since top quarks decay on timescales shorter than the one of QCD interactions,
these variables are sensitive to the $t\bar{t}$ spin correlation which was measured 
both at the Tevatron $p\bar{p}$~\cite{D0:2011kcb} and the LHC $pp$ colliders~\cite{ATLAS:2012ao, CMS:2013roq, ATLAS:2019zrq, CMS:2019nrx}.
Such measurements thus provide additional sensitivity to the presence
of new particles above the $\ttbar$ threshold.
The $t\bar{t}$ spin correlation also provides
information about the spin and the
CP properties of
the new particle~\cite{Bernreuther:1993hq,Bernreuther:1997gs,
Barger:2006hm,Frederix:2007gi,Carena:2016npr}.
Focusing on new spin-0 particles,
we consider two benchmark scenarios: 
(i) a 400~GeV ALP with a relative width of $2.5\%$ 
and 
(ii) an 800~GeV ALP with a relative width of $5\%$. 
Scenario (i) is motivated by a local $3.5\, \sigma$ excess observed 
by the CMS collaboration in the 400~GeV
mass region~\cite{CMS:2019pzc},
which has sparked some attention in the
literature~\cite{Arganda:2021yms,Biekotter:2021qbc,Connell:2023jqq}. 
No excess at this mass value has been reported in the latest ATLAS result~\cite{ATLAS:2024vxm}.
We will investigate to what extent 
an ALP with the same mass and width 
can be distinguished from a pseudoscalar Higgs boson 
depending on the effective ALP--gluon
coupling, even if both particles are produced with
the same total cross section.
We showcase that the LHC has
significant discovery potential
for ALPs in this mass range in the near future.
Under the assumption that no deviations from the
SM expectation will be observed,
we set current and projected limits
at several stages of the (HL-)LHC \program\
on the ALP couplings
to fermions and gluons in terms of the
Wilson coefficients of the linear
representation of the ALP--SM Lagrangian.
Furthermore, we 
compare these limits to the ones from 
existing experimental bounds from LHC searches in other final 
states, most notably from searches for resonances decaying into 
di-photons~\cite{ATLAS:2021uiz}
and into a $Z$ boson and a SM-like Higgs 
boson~\cite{ATLAS:2022enb}, and
measurements of the production
of four top-quarks~\cite{CMS:2023ftu}.
We also compare our bounds to other experimental limits, for instance from the
study of renormalisation group (RG) running
effects that mix ALP Effective Field Theory (EFT) operators and
dimension-six
Standard Model Effective Field Theory (SMEFT)
operators,
denoted ALP--SMEFT
interference~\cite{Galda:2021hbr,Biekotter:2023mpd}. 

The paper is organized as follows.
In \refse{sec:theory} we introduce the
theoretical framework, calculate the
partial width of the ALP and describe
the Monte Carlo simulation used in our analysis.
In \refse{sec:results} we present our main results, namely ALP bounds from existing searches, 
the analysis of the sensitivity for distinguishing
an ALP from a pseudoscalar Higgs boson, and projected ALP bounds at the LHC,
and we compare these to other existing limits.
We summarize our results and conclude
in \refse{sec:conclusion}.

\section{Theoretical framework and event simulation}
\label{sec:theory}

\subsection{The ALP Lagrangian}

ALPs are pseudoscalars which
preserve the softly broken axion shift symmetry $a(x) \to a(x) + c$, with $c$ being a constant. 
The general linear ALP--SM Lagrangian at dimension five~\cite{Georgi:1986df} is given by\footnote{The operator $\mathcal{O}_{a \phi} =
i (H^\dagger \overleftrightarrow{D}_\mu H ) \frac{\partial^\mu a}{f_a}$, where 
$H^\dagger \overleftrightarrow{D}_\mu H  = H^\dagger (D_\mu H) - (D_\mu H^\dagger) H$, is redundant as it can be rewritten in terms of the ones 
shown in \refeq{eq:lagrderiv}
by means of field redefinitions~\cite{Bauer:2020jbp}.}
\begin{align}
    \mathcal{L} \; = \; & \mathcal{L}_\textrm{SM} 
    + \frac{1}{2} (\partial_\mu a ) (\partial^\mu a ) +\frac{m_a^2}{2} a^2 - \frac{a}{f_a} c_{G} \, G^a_{\mu \nu} \Tilde{G}^{a \mu \nu} 
    - \frac{a}{f_a} c_{B} \, B_{\mu \nu} \Tilde{B}^{\mu \nu} 
    \nonumber
    \\
    & 
    - \frac{a}{f_a} c_{W} \, W^I_{\mu \nu} \Tilde{W}^{I \mu \nu} 
     - \frac{\partial^\mu a}{f_a} \sum_f \bar \Psi_f
      \mathbf{c}_f \gamma_\mu \Psi_f \, ,
    \label{eq:lagrderiv}
\end{align}
where $f_a$ and $m_a$ denote the ALP decay constant and mass,
respectively,
and $G$, $W$ and $B$ are the gauge fields associated 
to the $SU(3)$, $SU(2)$ and $U(1)$ gauge symmetries of the strong and electroweak interactions. 
The sum in the last term runs over the
fermionic fields $f = u_R, d_R, Q_L, L_L, e_R$.
In principle, the $\mathbf{c}_f$ are $3 \times 3$ matrices
in flavor space, but we neglect flavor mixing
in the following.
The invariance of the ALP couplings
under the transformation
$a \to a + c$
is manifest in the couplings of
the ALP to fermions, which are expressed
in terms of the derivative of~$a$.
Additional operators
arise from the transformation $a \to a + c$
when it is applied to the operators that couple the
ALP to the
gauge fields. These terms can be removed (besides
instanton effects in
QCD~\cite{Weinberg:1977ma,Wilczek:1977pj}) by
field redefinitions.
The ALP mass term softly breaks the axion
shift symmetry.
The presence of this explicit
breaking of the shift symmetry allows for
heavier ALPs compared to the classical QCD
axion whose mass is generated only by
non-perturbative QCD effects. Different
possibilities to generate the ALP mass
term in such a way 
that the possible ALP mass window
is extended
to larger masses (while maintaining a
solution to the strong QCD problem)
have been proposed in the literature~\cite{Holdom:1982ex,Rubakov:1997vp,Berezhiani:2000gh,Gianfagna:2004je,Hsu:2004mf,Hook:2014cda,Fukuda:2015ana,Chiang:2016eav,Dimopoulos:2016lvn,Gherghetta:2016fhp,Kobakhidze:2016rwh,Agrawal:2017ksf,Agrawal:2017evu,Gaillard:2018xgk,Hook:2019qoh,Csaki:2019vte,Co:2019jts,Gherghetta:2020ofz,Kivel:2022emq,Gavela:2023tzu},
e.g.~via additional strong 
interactions 
or via the axion kinematic misalignment mechanism.

The form of the ALP Lagrangian as shown in
\refeq{eq:lagrderiv} makes the shift symmetry
explicit in the ALP-fermion couplings.
For our analysis, it is more convenient to work in a basis which makes the connection of the ALP with 
a generic pseudoscalar, for instance in the 2HDM, more apparent. 
To this end, we re-write the fermionic operators
in terms of dimension-four Yukawa-like ALP--fermion couplings.
In this basis the effective Lagrangian can be
written as~\cite{Brivio:2017ije, Bauer:2020jbp, Bonnefoy:2022rik}
\begin{align}
  \mathcal{L} = \mathcal{L}_{\rm SM} +&
  \frac{1}{2} (\partial_\mu a ) (\partial^\mu a ) +
  \frac{m_a^2}{2} a^2 
  - \frac{a}{f_a} c_{\tilde G} \, G^a_{\mu \nu}
    \Tilde{G}^{a \mu \nu}
    - \frac{a}{f_a} c_{\tilde B} \, B_{\mu \nu}
    \Tilde{B}^{\mu \nu}
    - \frac{a}{f_a} c_{\tilde W} \, W^I_{\mu \nu}
    \Tilde{W}^{I \mu \nu}
    \notag \\
  &  +  \frac{a}{f_a}  \left[  \bar Q_L \tilde{H}   \,  {\bf \tilde Y}_U u_R 
  +   \bar Q_L H   \,  {\bf \tilde Y}_D d_R  
  +   \bar L_L  H  \,  {\bf \tilde Y}_E e_R 
  + \text{h.c.} \right] \, ,
  \label{eq:nonderiv}
\end{align}
where 
$\tilde {\bf Y}_f= i \left( {\bf Y_f} \, {\bf c_{f, R}} - {\bf c_{f, L}}\, {\bf Y_f} \right)$, with $\bf Y_f$ being the SM Yukawa couplings, and
$\tilde H = i \sigma_2 H$,
where $H$ is the SM Higgs doublet.
Furthermore, the fermion couplings are written as
${\bf c_{f,R}} = {\bf c_{u}}, \, {\bf c_{d}}, \, {\bf c_{e}}$ and ${\bf c_{f,L}} = {\bf c_{Q}}, \, {\bf c_{L}}$ for quarks and leptons. 
Note that the couplings of the axion to the gauge fields
in \refeq{eq:nonderiv}, written with a tilde (e.g.\ $\cG$), and those in the more manifestly
shift-invariant Lagrangian shown in \refeq{eq:lagrderiv} (e.g.\ $c_G$) are in general different (but related) parameters.

In our study, we only consider ALP couplings to top quarks and gluons, thus setting 
$c_{\tilde{W}} = c_{\tilde{B}} = 0$ and ${\bf c_d} =  {\bf c_L} = {\bf c_e} = 0$.
Rewriting the ALP couplings to up-type quarks yields
\begin{align}
\mathcal{L}_\textrm{up} 
    &= \frac{ia}{f_a}  \bar Q_L \tilde H \left(  {\bf Y_U} \, {\bf c_{u}} - {\bf c_{Q}} \, {\bf Y_U} \right) u_R + \textrm{h.c.} \nonumber \\
    &= \frac{ia}{f_a}  \bar q \tilde H \left( c_{u}^{33} - c_{Q}^{33}\right) Y_t \, t_R + \textrm{h.c.} \, + \,\dots \nonumber \\
    &= c_t \frac{ia}{f_a} \,  \bar q \tilde H \, Y_t \, t_R + \textrm{h.c.} \, + \,\dots \, ,
\end{align}
where $Y_t$ is the top-quark Yukawa coupling, 
we have defined $c_t \equiv c_{u}^{33} - c_{Q}^{33}$, and the ellipsis refers to terms involving first and second generation quarks. 
Moreover, $t_R$ is the right-handed top-quark spinor,
and the left-handed top- and bottom-quark
spinors are contained in the $SU(2)$ doublet
$q = (t_L , b_L)^T$.
With these simplifications, we finally obtain the following form of the ALP Lagrangian
\begin{equation}
    \mathcal{L} = \mathcal{L}_\text{SM} + \frac{1}{2} (\partial_\mu a ) (\partial^\mu a ) +\frac{m_a^2}{2} a^2 - \frac{a}{f_a} c_{\Tilde{G}} \, G^a_{\mu \nu} \Tilde{G}^{a \mu \nu}   
     + i c_{t} \, \frac{a}{f_a} \left( \bar q \,  Y_t \, \tilde H \, t_R + \mathrm{h.c.} \right) \, , 
    \label{eq:lagr}
\end{equation}
which we will use for our analysis below. 
This form of the Lagrangian facilitates the comparison with other models including pseudoscalars. 
One of the primary aims of 
the present paper is to investigate the potential to distinguish 
between an ALP with generic effective couplings to the top quark and 
gluon from a state
which only couples to gluons effectively via a top quark loop, i.e.\ 
for which $\cG =0$.
As already stated above, we denote this second scenario \textit{pseudoscalar Higgs boson} in order to distinguish it from the generic case. 
In order to facilitate the comparison with 
the CMS analysis of \citere{CMS:2019pzc}, we repeat here the considered Lagrangian (using the notation of~\citere{CMS:2019pzc}):
\begin{equation}
    \mathcal{L}_A =  \frac{1}{2} (\partial_\mu A ) (\partial^\mu A ) +\frac{m_A^2}{2} A^2  
    + i  g_{At\bar{t}} \, \frac{m_t}{v} \bar t  \gamma_5 \, t  \, A \, ,
    \label{eq:lagr_pseudoscalar}
\end{equation}
where $m_t = v Y_t / \sqrt{2}$ is the top-quark mass,
and $v \approx 246 \gev$ denotes the vacuum expectation value
of the Higgs field.
Comparing Eqs.~\eqref{eq:lagr} and~\eqref{eq:lagr_pseudoscalar} after EW symmetry breaking, 
we find that the two expressions are equal to each other for
\begin{equation}
    \cG = 0 \, , \qquad \ct = g_{A t \bar{t}} \, \frac{f_a}{v} \, .
    \label{eq:cplrelintro}
\end{equation}
Furthermore, in order to compare to recent work presented in the derivative basis~\cite{Blasi:2023hvb}, 
we note that for the considered case where the ALP couples only to
gluons and top quarks 
the gluon couplings in Eqs.~\eqref{eq:lagrderiv} and \eqref{eq:nonderiv} are related by~\cite{Bauer:2020jbp}
\begin{equation}
c_{\tilde G} = c_{G} + \frac{\alpha_s}{8 \pi} c_t \ ,
\label{eq:top-philic}
\end{equation}
where $\alpha_s$ denotes the QCD coupling.
In particular, a model in which the ALP couples exclusively to the top quark via a derivative coupling, $c_G =0$, corresponds to the case
$\cG = \alpha_s/(8 \pi) c_t$ in the notation adopted in our paper. We refer to this scenario as \textit{top-philic} to facilitate the comparison with~\citere{Blasi:2023hvb}. 

The total width of the ALP is treated as a free parameter in our analysis. 
In this way, we can account for the cases of possible ALP couplings to SM particles beyond top quarks and gluons leading to additional ALP decay channels and also of possible ALP decays to further beyond SM (BSM) particles, for instance decays to particles that are undetectable at the LHC. Both of these cases enter the $gg \to t \bar t$ process 
only via their effect of the total width and the corresponding modification of
the $a \to \ttbar$ branching ratio. Keeping the ALP width as a free parameter allows us to account for these possible additional ALP interactions.
In our analysis below, we will indicate the parameter regions where the sum of the partial widths of the ALP decays into $t \bar t$, $gg$ and $\gamma\gamma$ would be larger than the assumed total width.

\subsection{Effective ALP couplings}
\label{sec:eff_couplings}
In the following, we discuss the effective couplings of the ALP to gluons and photons, including effects from top-quark loops. 

The effective ALP--gluon--gluon vertex receives contributions
in our setup from the operator proportional to $c_{\tilde{G}}$
shown in \refeq{eq:lagr} and from the top-quark loop.
The effective $agg$ coupling is given by\footnote{We note that the constant shift $-1$ in \refeq{eq:gaggeff}
is present (similarly to the case of a CP-odd
Higgs boson) since we expressed the top-quark coupling
in the form of \refeq{eq:lagr_pseudoscalar}.
This constant piece is absent
if one instead uses the explicitly shift-invariant form
of the ALP--fermion operator.
Both formulations are
physically equivalent, since the
constant piece can
be absorbed via a linear shift of
the Wilson coefficients (see \citere{Bauer:2017ris}
for details).}
\begin{equation}
g_{agg}^\mathrm{eff} =  \frac{c_{\tilde{G}}}{f_a} +
  \frac{i}{2} \frac{\alpha_s}{4 \pi} \frac{c_t}{f_a}
  \left[ B_1\left( \frac{4 m_t^2}{m_a^2} \right) - 1 \right] \ ,
\label{eq:gaggeff}
\end{equation}
where the loop function is given
by $B_1(\tau) = 1 - \tau f^2(\tau)$ with
\begin{equation}
    f(\tau) = \frac{\pi}{2} + \frac{i}{2} \ln \left( \frac{1 + \sqrt{1-\tau}}{1 - \sqrt{1-\tau}} \right) \, .
\end{equation}

For the case of a non-vanishing
ALP--top-quark coupling,
the ALP obtains loop-induced ALP--photon--photon and ALP--$Z$--$h$ couplings.
While these couplings
enter
the $pp \to a \to \ttbar$
process mainly indirectly via
their impact on the $a \to t \bar t$ branching ratio, see above, they are furthermore relevant in this context
because they give rise to additional constraints
on the ALP parameter space from resonant di-photon  and $Zh$
searches
at the LHC.
The corresponding  $a\gamma\gamma$ 
and $aZh$ vertices
can be expressed in terms of the effective couplings
\begin{align}
    g_{a\gamma\gamma}^\mathrm{eff} &= 
      i \frac{\alpha}{4 \pi} N_c Q_t^2 \frac{c_t}{f_a} \left[ B_1\left( \frac{4 m_t^2}{m_a^2} \right) - 1 \right]  \ ,
      \label{eq:gayyeff} \\
    g_{aZh}^\text{eff} &=  \frac{N_c}{16 \pi^2} \frac{c_t}{f_a} \left(\frac{ \sqrt{2} m_t}{v} \right)^2 F \, ,
\end{align}
where $N_c = 3$ and $Q_t = 2 / 3$
are the \colour\ multiplicity and the
electric charge of the top quark, respectively,
$\alpha$ is the fine-structure constant, 
the loop function $B_1$ is identical to the one
for the ALP--gluon coupling given in
\refeq{eq:gaggeff},
and the integral $F$ can be approximated as $F = - \frac{m_t^2}{ m_a^2}  \left(\ln \frac{m_a^2}{m_t^2} - i \pi \right)^2 + \mathcal{O}\left(\frac{m_t^4}{m_a^4}\right)$ for 
$m_a > m_t$~\cite{Bauer:2016ydr}.\footnote{The contribution to $g_{a\gamma\gamma}^\mathrm{eff}$ from ALP--pion mixing is suppressed by factors of $m_{\pi^0}/m_a$ and hence not relevant here.}

\subsection{Partial widths of the ALP}
\label{sec:widths}
In the mass region we are investigating here, the most relevant
decay modes of the ALP are the decays into top-quark pairs
and into gluons, 
as well as
the loop-induced decays into photons,
see the discussion above, and the
loop-induced decay into a $Z$ boson
and the 125~GeV Higgs boson.
A discussion of additional decays of the
ALP that are generated at one-loop level
can be found in \refap{app:moredecays}.

The partial width for the decay into top quarks can be
written at leading order as
\begin{equation}
    \Gamma (a \to t \Bar{t} ) = \frac{m_a m_t^2 N_c |g_t^\text{eff}|^2}{8 \pi}
    \sqrt{1 - \frac{4 m_t^2}{m_a^2}} \, .
    \label{eq:widtt}
\end{equation}
We assume that sub-leading QCD corrections that would 
enter the effective top-quark coupling $g_t^\text{eff}$ 
are negligible.
In addition to the direct ALP--top--quark
coupling $c_t$,
contributions from diagrams involving
$\cG$ arise at loop-level.
These next-to-leading order effects are neglected in our analysis.
Thus, in our analysis we use
$g_t^{\rm eff} = c_{t} / f_a$
with $c_{t}$ defined in \refeq{eq:lagr}.

The partial width for the ALP decay into
gluons is given by~\cite{Bauer:2017ris}
\begin{equation}
    \Gamma(a\to gg ) = \frac{2 \, m_a^3  N_c^2 |g_{agg}^\text{eff}|^2}{9 \pi} \left[ 1+ \frac{83}{4}  \, \frac{\alpha_s(m_a)}{\pi} \right] \ ,
\end{equation}
where the second term in the brackets contains
the leading one-loop QCD 
corrections~\cite{Spira:1995rr},
and the effective ALP-gluon
coupling $g_{agg}^{\rm eff}$
is given in \refeq{eq:gaggeff}.

As explained above, we assume vanishing contact interactions
between the ALP and the weak gauge bosons,
i.e.~$c_{\tilde B} = c_{\tilde W} = 0$.
Thus, at leading order the decay of
the ALP to photons is induced only
through a top-quark loop.
The corresponding partial decay width can be written as
\begin{equation}
    \Gamma(a \to \gamma \gamma ) = \frac{m_a^3 |g_{a\gamma\gamma}^\text{eff}|^2}{4 \pi} \ ,
\end{equation}
where the effective coupling $g_{a\gamma\gamma}^{\rm eff}$
is given in \refeq{eq:gayyeff}.

In the considered mass range, the ALP can also decay to the SM-like Higgs bososn and a $Z$ boson via a top-quark loop. 
Assuming that the 125~GeV Higgs boson~$h$ is
purely CP-even as predicted by the~SM, 
the partial $a \to h Z$ decay rate 
is given by~\cite{Bauer:2016ydr,Bauer:2016zfj} 
\begin{align}
    \Gamma(a \to Z h)
    &=   \frac{m_a^3}{16 \pi} \, |g_{aZh}^\text{eff}|^2 \,\lambda^{3/2} \left(1, \frac{m_h^2}{m_a^2}, \frac{m_t^2}{m_a^2} \right) \, ,  
    \end{align}
$\lambda(x,y,z) = (x - y - z)^2 - 4yz$.

In the next sections, we will consider the impact of an ALP on $\ttbar$ production at the LHC.
As already discussed above, potential additional ALP decays which would modify 
the $a \to \ttbar$ branching ratio
are taken into account by keeping the total ALP width as a free parameter. 
This includes additional top-quark-loop induced contributions, 
e.g.\ into the electroweak gauge bosons, decays induced through effective ALP--SM
couplings beyond the coupling to gluons and top quarks, and ALP decays into additional 
BSM particles, see also
\refap{app:moredecays}.
For the considered benchmark scenarios below, the branching ratio into 
$t \bar t$ typically 
dominates. For instance, for an ALP
at $\ma = 400$~GeV with $c_t/f_a =3.0\tev^{-1}$,
$c_{\tilde{G}}/f_a=0.015\tev^{-1}$, a fixed width of $\Gamma/m_a = 2.5\%$, and using the top-quark pole mass of $m_t = 172.5$ GeV,
the branching ratios for the decays
into SM particles considered in this analysis are
$\textrm{BR}(a \to t \Bar{t}) = 65 \,\%$,
$\textrm{BR}(a \to gg) = 0.84\,\%$, 
 $\textrm{BR}(a \to Z h) = 0.13\,\%$
and
$\textrm{BR}(a \to \gamma \gamma) =
1.3 \cdot 10^{-5}$.

\subsection{Monte Carlo simulation setup and observables}
\label{sec:vegas}

\begin{figure}
    \centering
    \begin{fmffile}{toploop}
     \begin{fmfgraph*}(140,80)
       \fmfleft{i1,i2}
       \fmfright{o1,o2}
       \fmf{gluon,tension=0.6}{i1,l1}
       \fmf{gluon,tension=0.6}{i2,l2}
       \fmf{phantom,tension=0.3}{l1,o1}
       \fmf{phantom,tension=0.3}{l2,o2}
       \fmflabel{$g$}{i1}
       \fmflabel{$g$}{i2}
       \fmffreeze
       \fmf{fermion,tension=.5,label=$t$,label.side=right}{l2,l1}
       \fmf{fermion,tension=.5}{l1,v1}
       \fmf{fermion,tension=.5}{v1,l2}
       \fmf{fermion,label=$\bar{t}$,tension=.5,label.side=right}{o1,v2}
       \fmf{fermion,label=$t$,tension=.5}{v2,o2}
       \fmf{dashes,label=$a$,tension=.5}{v1,v2}
       \fmfdot{v1}
       \fmfdot{v2}
     \end{fmfgraph*}
    \end{fmffile}
    \quad
    \begin{fmffile}{gluoncoupling}
     \begin{fmfgraph*}(140,80)
       \fmfleft{i1,i2}
       \fmfright{o1,o2}
       \fmf{gluon}{v1,i1}
       \fmf{gluon}{v1,i2}
                     \fmflabel{$g$}{i1}
       \fmflabel{$g$}{i2}
       \fmf{fermion,label=$\bar{t}$}{o1,v2}
       \fmf{fermion,label=$t$, label.side=right}{v2,o2}
       \fmf{dashes,label=$a$}{v1,v2}
       \fmfdot{v1}
       \fmfdot{v2}
     \end{fmfgraph*}
    \end{fmffile} 
    \\[2mm]
    \caption{BSM Feynman diagrams contributing to the process $gg \to \ttbar$
    (for simplicity, the decay of the produced top quarks, which is taken into account in our analysis, is not shown). The left diagram contains a top quark loop and scales with the coupling $\ct^2$, while the right diagram contains an additional effective
    $agg$ coupling and scales with $\cG \ct$.}
    \label{fig:feynman}
\end{figure}
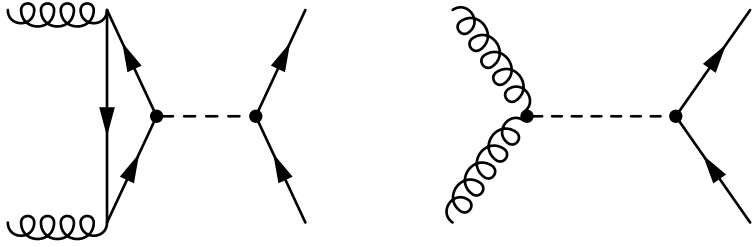

In view of the above discussion of the ALP couplings,
we consider two possible
BSM diagrams for the process $gg \to \ttbar$ which can be seen in \reffi{fig:feynman}
(where the decays of $t$ and $\bar t$ 
are omitted for simplicity): one containing a top-quark loop (left), scaling with the coupling $\ct^2$, 
and one containing the effective tree-level $agg$ coupling (right), scaling with $\cG \ct$. Both diagrams, as well as their interference with each other and with 
the SM background for $\ttbar$ production, contribute to the $a \to \ttbar$ signal, which in general depends non-linearly on both couplings $\ct$ and $\cG$.
In the absence of CP violation, as we assume throughout this paper, there is no interference contribution
between the $s$-channel exchange of the Higgs
boson at 125~GeV and the one of CP-odd BSM particles
(here ALP $a$ or pseudoscalar Higgs boson $A$).
Therefore, the Higgs boson at 125~GeV does not contribute to the signal in our analysis.

In the following we will
investigate the sensitivity
of LHC searches in the $t\bar t$ final state
to ALPs and we will \analyze\
differences between an ALP $a$ with $\cG \neq 0$ and a
pseudoscalar Higgs boson $A$
without additional contributions to
the gluon coupling besides the one
from the top-quark loop.
To this end,
we generate Monte Carlo~(MC) events of the
process $gg \rightarrow 
a/A \rightarrow \ttbar
\rightarrow {b \bar{b}
\ell^+ \ell^- \nu \bar{\nu}}$
at leading order (LO) in QCD using the
general-purpose MC generator
\textsc{MadGraph 5}~\cite{Alwall:2014hca}.
For the ALP events, we use an adapted version of the Universal FeynRules
Output (UFO) model~\cite{Degrande:2011ua}
provided in \citere{Brivio:2017ije},
which we modified to explicitly include
the quark loop-induced production using a
form factor taken from
\citere{Bonilla:2021ufe}.
For the pseudoscalar Higgs boson,
an in-house UFO model is used.

Events for the SM $\ttbar$ background are generated at next-to-leading order (NLO) in QCD using the MC generator \textsc{Powheg}  \cite{Nason:2004rx,Frixione:2007vw,Alioli:2010xd,Frixione:2007nw}. 
The NNPDF 3.1 parton distribution function (PDF) set~\cite{NNPDF:2017mvq} is employed for the generation of both the signals and the SM background.
The events are showered and hadronized with the \textsc{Pythia 8.3} program~\cite{Bierlich:2022pfr}.

To estimate higher-order effects on the 
event yields, we calculate the cross section for resonant $gg \rightarrow A$ production at NNLO in QCD
for a 2HDM pseudoscalar Higgs boson using the \textsc{2HDMC}~\cite{Eriksson:2009ws}
and \textsc{SusHi}~\cite{Harlander:2012pb} programs. 
We then define a $K$~factor $K_{\mathrm{res}}$ for the resonant $A$ signal as the ratio of the NNLO cross section to the LO one predicted by \textsc{MadGraph}. 
For the $A$/SM interference signal, we define the $K$~factor as $K_{\mathrm{int}} = \sqrt{K_{\mathrm{res}} K_{\mathrm{SM}}}$, where $K_{\mathrm{SM}}$ is the SM $K$~factor, which normalises to the NNLO+NNLL SM $\ttbar$ cross section of 833.9 pb as calculated with \textsc{Top++ 2.0}~\cite{Czakon:2011xx}.\footnote{More precise calculations are available only for a CP-even Higgs boson~\cite{Banfi:2023udd}.} The same $K$~factors 
are used for both the pseudoscalar Higgs boson as well as the ALP with $\cG \neq 0$.

\begin{figure}
\centering
\includegraphics[width=0.49\textwidth]{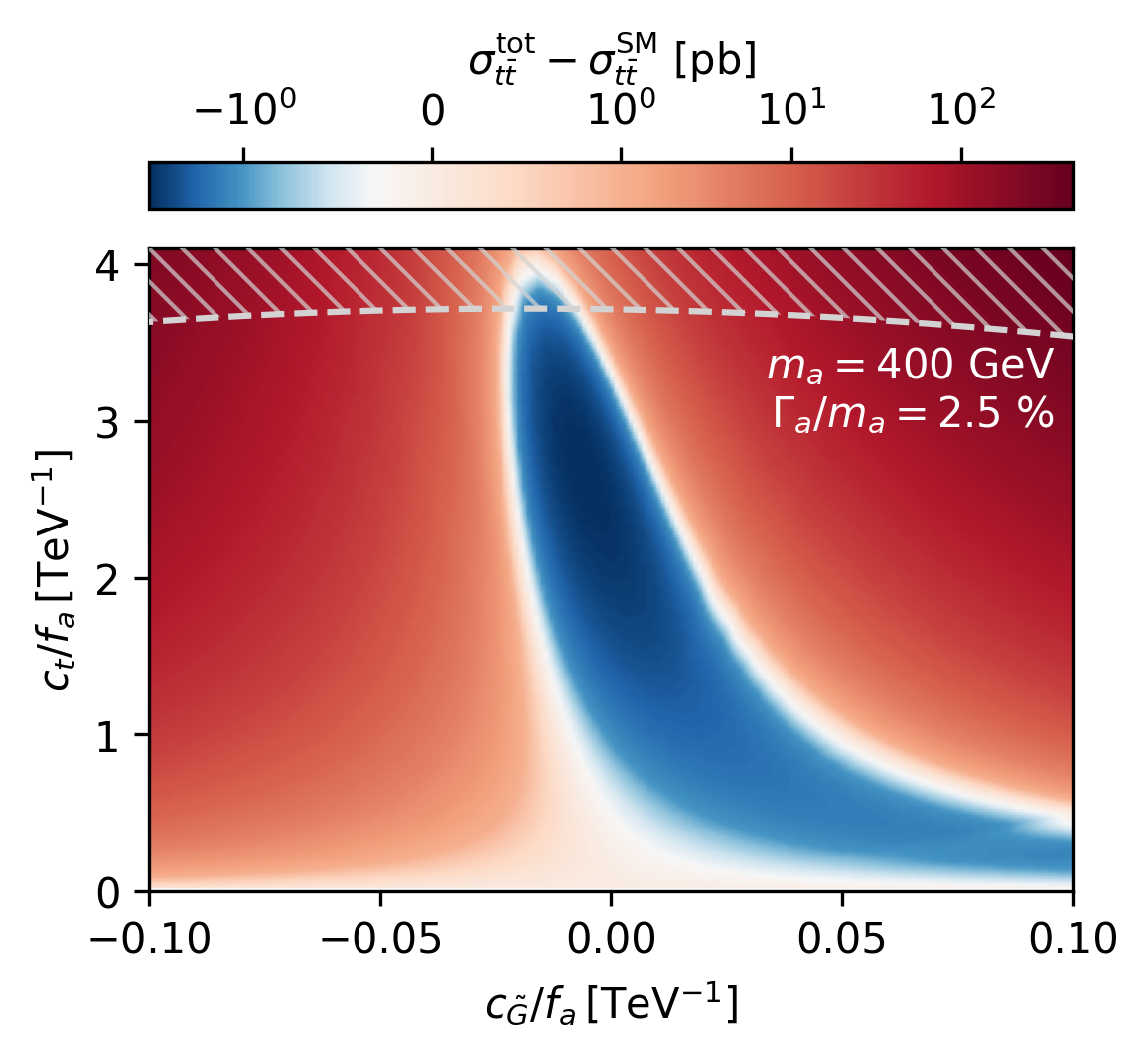}
\hfill
\includegraphics[width=0.49\textwidth]{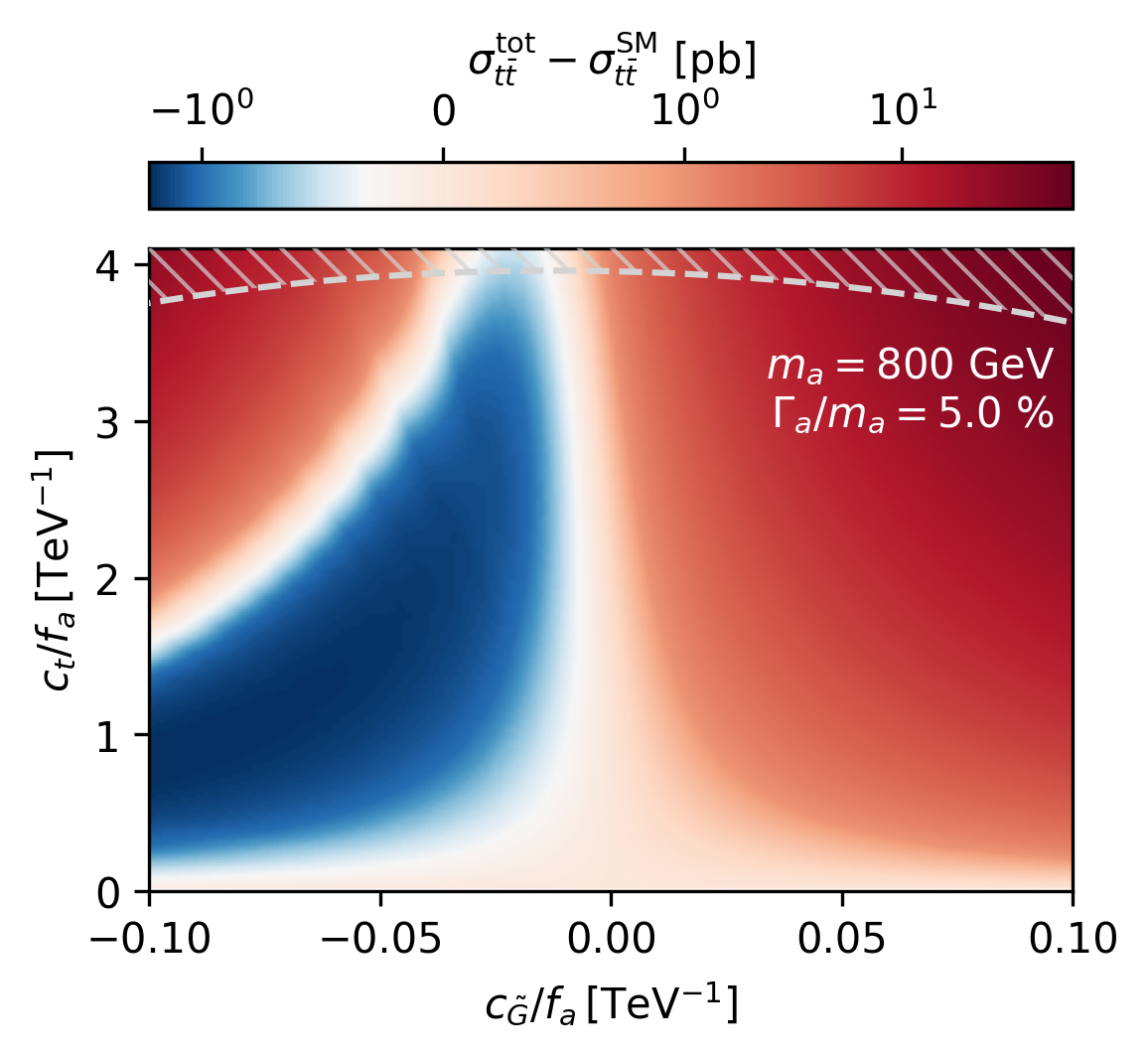}
\caption{The background-subtracted
inclusive $pp \to \ttbar$ cross section,
including resonant $pp \to a$ production and ALP/SM interference, 
$\sigma_{t \bar t}^{\rm tot} - \sigma_{t \bar t}^{\rm SM}$, is shown in the plane
of the ALP couplings $\cG/f_a$ and $\ct/f_a$ for different ALP masses and relative widths of $400 \gev$ and 2.5\% (left) as well as $800 \gev$ and 5.0\% (right). 
The hatched band shows the region in which the sum of the predicted
partial widths for the $a \to \ttbar,
gg, Zh, \gamma\gamma$ decays 
exceeds the assumed total width.
Negative values of $\ct$ are not shown as the cross section is symmetric under the sign change of both $\ct$ and $\cG$, see Eq.~\eqref{eq:xSec_para}.
}
\label{fig:crosssections}
\end{figure}

The results for the inclusive $\ttbar$ cross section 
incorporating the contributions from the resonant ALP signal 
and the ALP/SM interference, 
while the cross section corresponding to the SM $\ttbar$ 
background is subtracted, can be seen in \reffi{fig:crosssections} for the two ALP masses and widths of $\ma = 400 \gev$, $\ga / \ma = 2.5$\%
(left) and $\ma = 800 \gev$, $\ga / \ma = 5.0$\% (right). 
The background-subtracted result
$\sigma_{t \bar t}^{\rm tot} - \sigma_{t \bar t}^{\rm SM}$ is seen to be negative in a part of the $(c_{\tilde G}/f_a, c_t/f_a)$
parameter plane. This is
due to the destructive
interference between the diagrams shown
in \reffi{fig:feynman} and the SM diagrams contributing
to $t \bar t$ production.

Explicitly, the 
background-subtracted inclusive cross section 
for the $400 \gev$, 2.5\% case 
can be parameterized as
\begin{align}
   \sigma_{t \bar t}^{\rm tot} - \sigma_{t \bar t}^{\rm SM}  = &
    \Bigg\{ \left[ 0.109  \ct^4 + 22.3 \ct^3 \cG + 1960 \ct^2 \cG^2 \right] \left( \frac{\text{TeV}}{f_a} \right)^4 \nonumber \\
    &- \left[  1.20 \ct^2 + 97.2 \ct \cG  \right] \left( \frac{\text{TeV}}{f_a} \right)^2 
    \Bigg\} \text{ pb}\, .
    \label{eq:xSec_para}
\end{align}
Note the symmetry of the cross section under the sign change of both $\ct$ and $\cG$.

Following the CMS search for a heavy pseudoscalar Higgs boson~\cite{CMS:2019pzc}, we discriminate the signal and background events  based on two variables, the invariant mass of the $\ttbar$ system, $\mtt$, and the spin correlation variable $\chel$. The latter is defined as
\begin{equation}
    \chel = \cos{\varphi} = \hat{\ell}^{+} \cdot \hat{\ell}^{-}~,
\end{equation}
where $\varphi$ denotes the angle between the directions of flight $\hat{\ell}^{+}$ and $\hat{\ell}^{-}$ of the two leptons, defined respectively in the rest frames of their parent top or anti-top quarks.
It can be shown that the distribution of this variable (without phase space cuts) has the
form~\cite{Bernreuther:2004jv}
\begin{equation}
    \frac{1}{\sigma} \frac{d \sigma}{d \chel} = \frac{1}{2} \left( 1 - D \, \chel \right) \ ,
\end{equation}
where the slope $D$ is sensitive to the parity of a possible intermediate particle (in this case, the ALP or 
a pseudoscalar Higgs boson).\footnote{It should be noted
that our variable $\chel$ is called $\cos \varphi$ in \citere{Bernreuther:2004jv} and does not correspond to their variable $\chel$.} Thus, this observable can be used to 
discriminate between
the signal and the SM background.

\begin{figure}[t]
    \centering
    \includegraphics[width=0.95\textwidth]{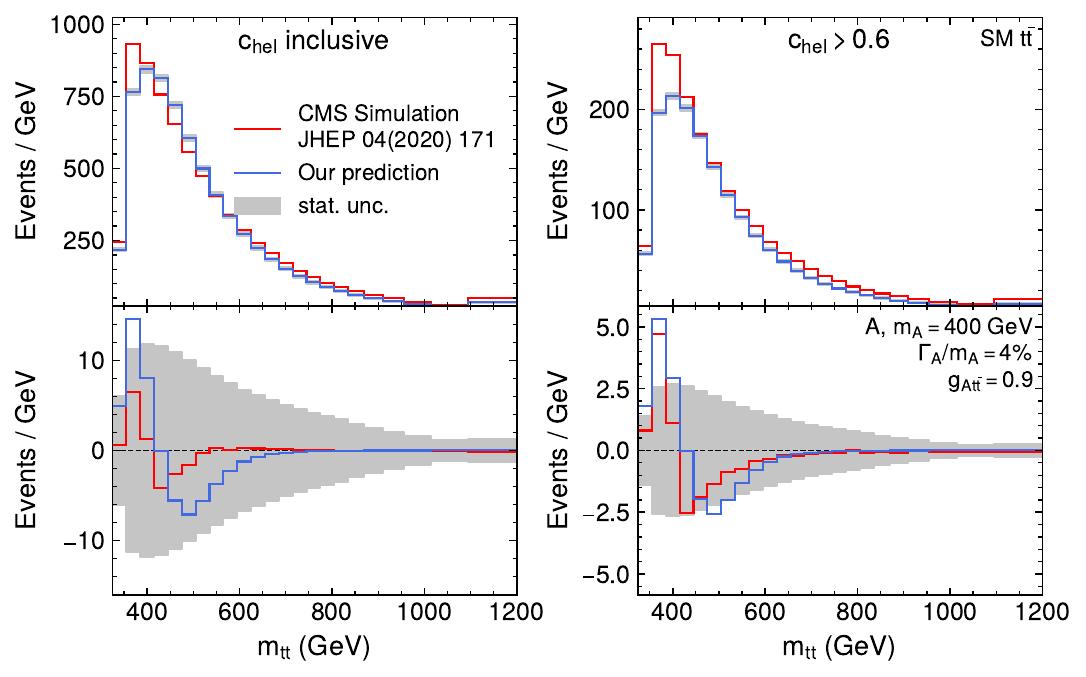}
    \caption{Differential distribution in $\mtt$ for the SM (top) and for a pseudoscalar Higgs boson with $m_A = 400 \gev$, 4.0\% width and a coupling strength of $\gAtt = 0.9$, 
    where the SM background has been subtracted
    (bottom). The left plot shows the distribution inclusive in the variable $\chel$, while the right plot shows it after selecting only events with $\chel > 0.6$. Our smeared prediction (blue) is compared to the CMS simulation taken from \cite{CMS:2019pzc} (red). All predictions are shown for an integrated luminosity of $35.9 \, \fb$. The gray bands show the expected statistical uncertainty from the SM $\ttbar$ background.
    }
    \label{fig:smearing}
\end{figure}

In order to 
account for the finite detector resolution
we apply a Gaussian smearing with a standard deviation of $\sigma = 15 \%$ 
on $\mtt$ .
The magnitude of the smearing was 
extracted from a fit of the smeared generator predictions 
to both the SM $\ttbar$ background and the pseudoscalar Higgs-boson signal after the full detector simulation in \citere{CMS:2019pzc}. 
This resolution is larger than all values of the relative ALP width considered in this analysis, and as such we expect the shape of the ALP signal to be mostly insensitive to $\Gamma_a$.
A comparison between our $\mtt$ distribution prediction and the CMS simulation is shown in \reffi{fig:smearing}. 
Note that we perform this comparison for $\Gamma_a/m_a = 4\%$ as distributions for lower values of the relative width are not displayed in~\citere{CMS:2019pzc}.
In the left panel, we show the distribution inclusive in $\chel$. In the right panel, we show the distribution after 
the cut $\chel > 0.6$, highlighting the discrimination power of this variable. We will employ this cut for the rest of our analysis.
While some discrepancy
for the signal can be seen for the case of the (less sensitive)
$\chel$-inclusive 
prediction, the distributions 
agree rather well with each other for $\chel > 0.6$.
For the SM background, some differences are present just above the $\ttbar$ threshold, which are expected to result from the details of the $\ttbar$ reconstruction in the experimental analysis.

We further approximate the experimental acceptance and efficiency for both signal and $\ttbar$ background as 10.6\% before the $\chel$ requirement based on the numbers reported by CMS~\cite{CMS:2019pzc}. This acceptance is defined as the fraction of $\ttbar \rightarrow b \bar{b} \ell^+ \ell^- \nu \bar{\nu}$ events ($\ell$ being electrons, muons or leptonically decaying taus) that pass all triggers and analysis cuts and contribute to the likelihood fit.

\subsection{Systematic uncertainties}
\label{sec:syst_uncert}

In analyzing the discrimination
between a BSM signal and the SM
expectation,
we consider the following sources of systematic uncertainties:
\begin{itemize}
    \item 
        Unknown higher-order corrections in the calculation of both the signal and the $\ttbar$ background. In both cases, the corresponding uncertainties are estimated by varying the renormalization and factorization scales independently up and down by a factor of 2. 
    \item The uncertainty in the PDF choice is estimated as the envelope of 100 pseudo-Hessian NNPDF 3.1 replicas, as recommended in \citere{Butterworth:2015oua}.
    \item  The value of the top-quark mass assumed in the simulation of the SM $\ttbar$ background. It is set to $m_t = 172.5$ GeV by default and assigned a Gaussian uncertainty of $1 \gev$, as in \citere{CMS:2019pzc}.
        \item The uncertainty of the
    total rate of the SM $\ttbar$ background. It is taken as a log-normal uncertainty of 6\% as in \citere{CMS:2019pzc}. The inclusion of this uncertainty does not significantly influence our results.
\end{itemize}

\begin{figure}
    \centering
    \includegraphics[width=0.52\textwidth]{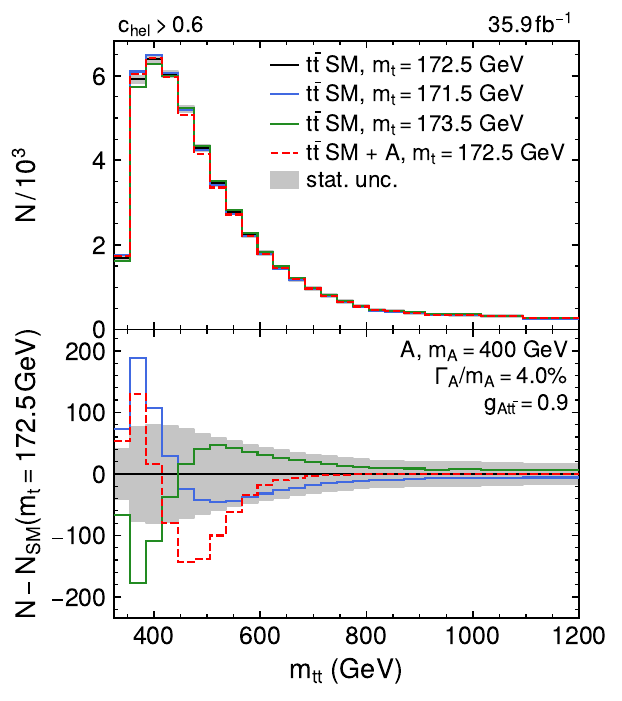}
    \hfill
    \includegraphics[width=0.47\textwidth]{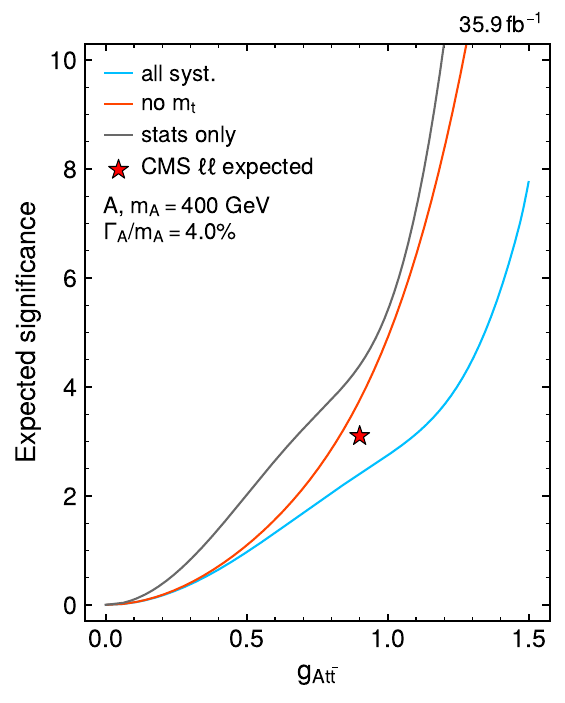}
    \caption{Left: The effect of a top-quark mass variation of $\pm 1 \gev$ in the SM $\ttbar$ background (green and blue lines; the central value is indicated by the black line in the upper plot, all displayed SM curves are normalized such that they yield the same total cross section) compared to the effect of a pseudoscalar Higgs boson with $m_A = 400 \gev$, 4\% width and $\gAtt = 0.9$ 
        (red dashed line). The gray band in the lower plot, where the SM background has been subtracted, indicates 
    the statistical uncertainty for an integrated luminosity of $35.9 \, \fb$.
    Right: The expected significance for the pseudoscalar Higgs boson as a function of its coupling $\gAtt$ for 
    the full set of systematic uncertainties, the full set except for the top-quark mass uncertainty, and for the case where only statistical uncertainties are taken into account. The expected significance reported by CMS for $\gAtt = 0.9$ is shown as the red star.} 
    \label{fig:systs}
\end{figure}

Among these, the top-quark mass uncertainty is of particular importance 
for ALPs or pseudoscalar Higgs bosons with masses close to the $\ttbar$ production threshold, where
the $\mtt$ distribution for the SM $\ttbar$ background is strongly affected by even small variations in the top-quark mass for low $\mtt$ values. In \reffi{fig:systs} on the left, we show the effect of such a variation for a luminosity of 35.9 $\fb$, and compare it to the effect of a pseudoscalar Higgs-boson signal corresponding to the
excess observed by CMS in the first-year Run 2
analysis~\cite{CMS:2019pzc}.
As expected, we find that the top-quark mass variation has significant
effects on the bins close to the $\ttbar$ threshold.
The comparison with the expected signal for a pseudoscalar Higgs boson at 400~GeV shows that the variation of the top-quark mass in the SM background by 
$-1 \gev$ yields, after subtracting the SM background with $m_t = 172.5 \gev$,
some similarity with the peak--dip structure that is expected for the signal.

Since the effects of experimental cuts are taken into account only using acceptance factors, variations in acceptance due to the top-quark mass dependence are not included in our analysis. In particular, lowering the top-quark mass will result in lower transverse momenta of the top-quark decay products (leptons and jets), which in an experimental analysis would result in more events being rejected by triggers and lepton or jet quality cuts. This in return would mitigate the steep increase of observed events for low $\mtt$ shown as the blue line in \reffi{fig:systs} (left). 
Similarly, the opposite is true for raising the
top-quark mass (green line), in total leading to a smaller uncertainty due to the top quark mass. 
In addition to this, our method imposes the requirement $\chel > 0.6$, while the experimental analysis considers the full range in $\chel$, split into five bins. 
A pseudoscalar Higgs boson signal is expected to contribute mostly for high $\chel$, while a variation due to a shift in the top mass affects all $\chel$ bins similarly, which gives additional power to distinguish the signal from a variation in the top-quark mass. As a result of both these effects,
the uncertainty due to the top-quark mass is likely overestimated in our setup, and we will consider our results both including and excluding the
uncertainty stemming from
the top-quark mass
in the following.

In order to compute expected significances and limits including the systematic uncertainties,
we perform hypothesis tests based on a binned profile likelihood fit 
with the package 
\texttt{pyhf}~\cite{pyhf,pyhf_joss}.
The expected number of events (SM background, resonant ALP production, and ALP--SM interference) in each bin of the differential distribution in $\mtt$, as shown in \reffi{fig:smearing}, can be parameterized as a polynomial in the two ALP couplings $\ct / f_a$ and $\cG / f_a$. With this, we define the likelihood
\begin{equation}
\label{eq:likelihood}
    \mathcal{L}(\ct, \cG, \theta_j) = \prod_i \mathrm{Poisson} (N^{\mathrm{obs}}_i | N^{\mathrm{pred}}_i (\ct, \cG, \theta_j)) \times \prod_j p(\theta_j) \ ,
\end{equation}
where
$N^{\mathrm{obs}}_i$ is the observed number of events in bin $i$, $N^{\mathrm{pred}}_i$ is the predicted number of events for given values of the couplings, and $\theta_j$ are nuisance parameters encoding different theory-based systematic uncertainties as discussed above along with their corresponding prior distributions 
$p(\theta_j)$. These are given by log-normal (for the $\ttbar$ rate uncertainty) or Gaussian (for all other uncertainties) distributions, with standard deviations as given above.
Both shape and rate effects of the different uncertainty sources $\theta_j$ are fully taken into account in the predicted number of events $N^{\mathrm{pred}}_i(\ct, \cG, \theta_j)$.
For the HL-LHC projection all systematic uncertainties are halved since the accuracy of the theoretical 
predictions is expected to improve significantly on the relevant timescales. 
In the fit, the likelihood $\mathcal{L}$ is optimized simultaneously as a function of the couplings $\ct/f_a, \, \cG/f_a$ and the nuisance parameters $\theta_j$.

In order to derive an expected limit for the ALP couplings, we define a test statistic $t_{\ct, \cG}$~\cite{Cowan:2010js} as the profile likelihood ratio
\begin{equation}
\label{eq:teststat}
    t_{\ct, \cG} = \min_{\theta_j} \left( -2 \ln \frac{ \mathcal{L}(\ct, \cG, \theta_j) } { \hat{\mathcal{L}} } \right) 
    \ ,
\end{equation}
where
\begin{equation}
    \hat{\mathcal{L}} = \max_{\hat{c}_t, \hat{c}_{\Tilde{G}},  \hat{\theta}_j} \mathcal{L}(\hat{c}_t, \hat{c}_{\Tilde{G}}, \hat{\theta}_j) 
\end{equation}
is the value of the likelihood at the best-fit coupling and nuisance parameter values for given observed data. With this setup, the test statistic $t_{\ct, \cG}$ is a measure of the agreement between the observed data and the ALP prediction for given couplings $\ct$ and $\cG$, taking into account systematic uncertainties. 

To compute the expected significance
for the detection of an
ALP signal for given values of the ALP couplings, we assume that the observed data is equal to the prediction of 
the sum of the ALP signal and the SM $\ttbar$ background, and perform a hypothesis test for the background-only hypothesis, i.e.\ for the test statistic $t_{0,0}$ as defined in \refeq{eq:teststat}. The significance for rejecting the background-only hypothesis is given by $\sqrt{t_{0,0}}$ in this case.

We show
the expected significance at a luminosity of $35.9\,\fb$ for a pseudoscalar Higgs boson with $m_A = 400 \gev$, $4.0 \%$ width and varying coupling modifiers $\gAtt$ (related to $\ct/f_a$ via \refeq{eq:cplrelintro}) in \reffi{fig:systs} on the right for different uncertainty models: with all systematic uncertainties including the one stemming from
the top-quark mass, 
excluding the
one from the top-quark mass,
and with statistical uncertainties only.
For $\gAtt = 0.9$, corresponding 
to the CMS excess, we find an expected significance of 2.3 standard deviations 
if all uncertainties including the one from the
top-quark mass are taken into account,
and 3.7 
standard deviations for the case where the systematic uncertainty arising from the 
top-quark mass is not included.
Comparing this to the expected significance in the di-lepton channel reported by CMS of 3.1 
standard deviations~\cite{CMS:2019pzc},
we find that the value that was obtained in the experimental analysis
lies between our 
estimates when including or excluding the top-quark mass uncertainty. This is in line with our expectation,
as mentioned above, that we overestimate the effect of top-quark mass variations because we do not incorporate
acceptance effects. For the projected limits and significances that we will present below
we will always quote the significances both including and excluding this uncertainty.

We also compute the significances for distinguishing a general ALP $a$ from a pseudoscalar Higgs boson $A$ with couplings $\ct / f_a = \gAtt / v$ and $\cG = 0$ (see \refeq{eq:cplrelintro}). Similarly to the definitions above (\refeq{eq:teststat}), 
we assume the observed data 
to be equal to the expectation for the ALP, and calculate the expected significance for rejecting the pseudoscalar Higgs boson hypothesis as $\sqrt{t_{\ct,\cG=0}}$.

\section{Results}
\label{sec:results}

In this section we present the resulting
limits on the 
Wilson coefficients of the ALP Lagrangian given in \refeq{eq:lagr}
and \analyze\ the sensitivity for
distinguishing between an ALP with
non-vanishing $\cG$ coupling and a pseudoscalar
Higgs boson of an extended Higgs sector 
for which $\cG = 0$.

\subsection{Translation of pseudoscalar Higgs boson limits}

\begin{figure} 
\centering
\includegraphics[width=0.48\textwidth]{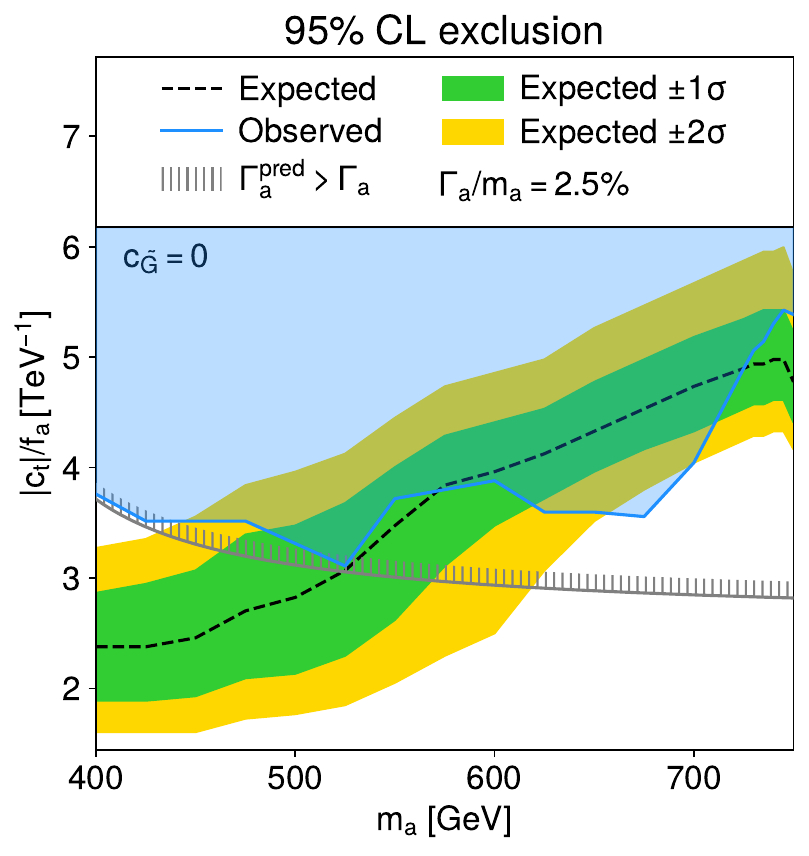}
\hfill
\includegraphics[width=0.48\textwidth]{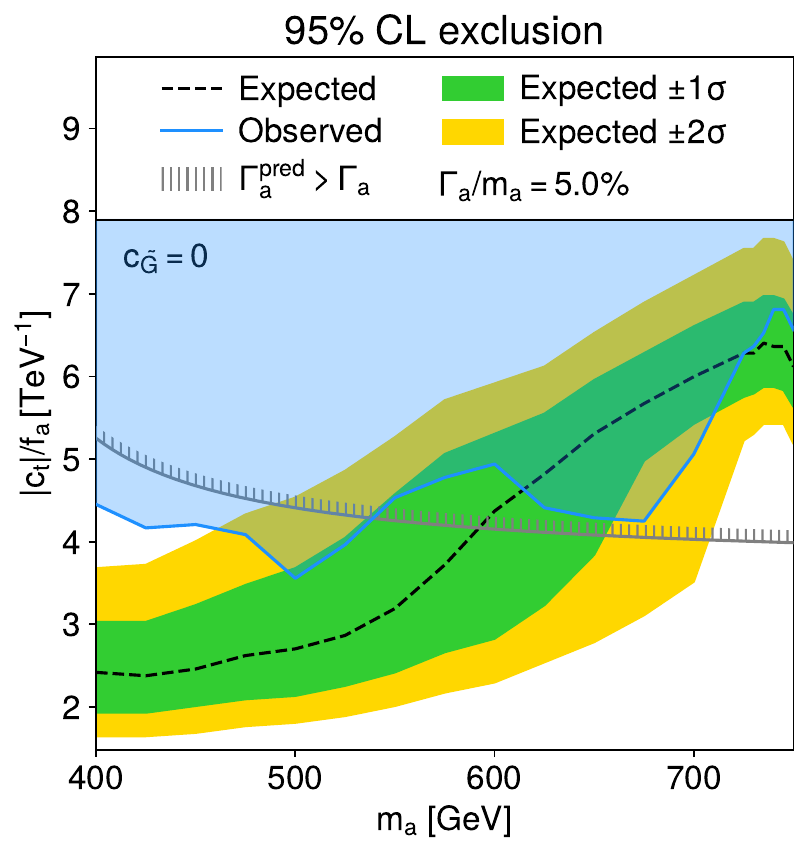}
\caption{Limit on the coupling of an ALP to the top quark, $\ct/f_a$, in the case $\cG = 0$, translated from 
\citere{CMS:2019pzc}, for a relative ALP width of 2.5\% (left) and 5\% (right). The hatched band shows the region in which the sum of the predicted
partial widths for the $a \to \ttbar,
gg, Zh, \gamma\gamma$ decays 
exceeds the assumed total width.
}
\label{fig:cmslimits}
\end{figure}

Since the additional pseudoscalar Higgs boson considered in \citere{CMS:2019pzc} and an ALP exhibit the same coupling structure for $c_{\tilde{G}} = 0$, 
the existing limits on the process $pp \rightarrow A \rightarrow t \bar{t}$ can be directly translated 
in this case
into the first experimentally observed upper limits on the ALP coupling to the top quark using \refeq{eq:cplrelintro}.
In \reffi{fig:cmslimits} 
we show the expected (black dashed) and observed (blue)
upper limits on $\ct/f_a$
as a function of the ALP mass $\ma$ assuming a total
relative width of $\ga / \ma = 2.5$\%
and $\ga / \ma = 5$\% in the
left and the right plot, respectively,
based on the results of the CMS
search for additional Higgs bosons in~$t\bar{t}$ final
states  using $35.9~\fb$ of data~\cite{CMS:2019pzc}.
Also shown are the $1\sigma$ and $2\sigma$
uncertainty bands of the expected limits
as green and yellow bands, respectively.
Coupling values 
for which the predicted
total width of the ALP, taking into account
the $a \to \ttbar$, 
$a \to gg$ and $a \to \gamma\gamma$ decays, is larger than the
assumed total width are indicated by the gray 
hatched band.

In the left plot of \reffi{fig:cmslimits} we
can observe that in the ALP mass region
$400\gev < m_{a} \lesssim 550\gev$
upper limits of $|\ct| / f_a$ between 3.0 and 3.8 $\tevinv$ are found.
The expected limit in this mass region
of $|\ct| / f_a \lesssim 2.5\tevinv$ is substantially
smaller than the observed limits.
This is a manifestation of the local excess
observed by CMS (which, however, is not supported by the ATLAS result~\cite{ATLAS:2024vxm}). 
For larger masses the expected and observed limits are located in the parameter region where the
predicted total width
is larger than the assumed total width, 
so that no limits on $|\ct|/f_a$ that are compatible with the assumption of a 2.5\% total width can be inferred.
In the right plot a relative ALP
width of 5\% is assumed.
In
the mass range $400\gev < m_{a} \lesssim
550\gev$, where both the expected and the observed
limit lie below the hatched band 
(and therefore the obtained results are compatible with the 
assumption of a 5\% total width)
coupling values of $|\ct| / f_a \gtrsim 4\tevinv$ are excluded.

\subsection{Discrimination between an ALP and a 2HDM pseudoscalar Higgs boson}
\label{sec:unterschied}

We now consider the case of an additional effective
ALP--gluon coupling, $\cG \neq 0$, and investigate the sensitivity for distinguishing an ALP from
a pseudoscalar Higgs boson with 
the same mass and total width
and for which we assume $\cG = 0$. 
Using the MC simulation described in \refse{sec:vegas}, we 
\analyze\ the resulting differences in the $\mtt$ distribution.

\begin{figure} 
\centering
\includegraphics[width=\textwidth]{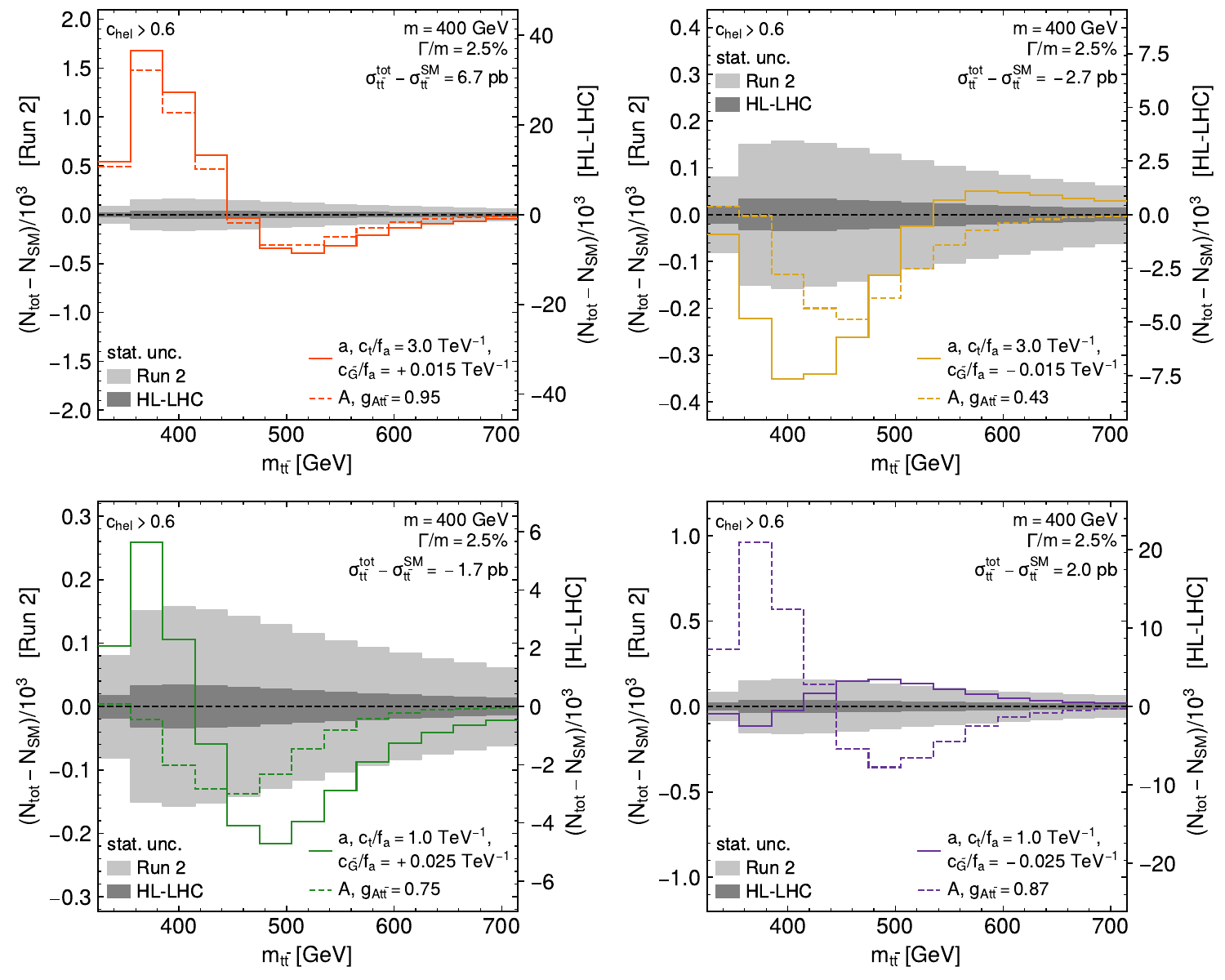}
\caption{Differential distribution in $\mtt$ for an ALP with different values of $\cG$ and $\ct$ and a 
pseudoscalar Higgs boson with different values of $\gAtt$,
both with a mass of $400\gev$ and a total width of 2.5\%. The couplings $\cG$, $\ct$ and $\gAtt$ are chosen in the considered benchmark scenarios
such that the 
ALP and the pseudoscalar Higgs boson have the same integrated cross section in a given panel.
Event counts are shown for integrated luminosities corresponding to Run 2 ($138 \, \fb$, left axis) and the HL-LHC ($3 \, \ab$, right axis). The gray
bands indicate the expected statistical uncertainties on the SM background for the two integrated luminosities. 
}
\label{fig:distributions}
\end{figure}

We show in Fig.~\ref{fig:distributions} the $\mtt$ distributions after the cut $\chel > 0.6$
for a pseudoscalar Higgs boson $A$
and for an ALP $a$ 
both with a mass of $400\gev$, for several benchmark values of $\cG/f_a$ and $\ct/f_a$, given in \refta{tab:benchmarks}.
In each plot panel, the coupling $g_{A\ttbar}$ of the pseudoscalar Higgs boson is chosen such that its total cross-section contribution matches the one
of the ALP for each benchmark.\footnote{For
$\ct =3 , \, \cG =-0.015$, both $g_{A\ttbar}=0.43$ and $g_{A\ttbar}=0.69$ lead to the same cross section. We have plotted the distribution for the coupling for which the ALP and pseudoscalar Higgs lines are closer to each other.}
Two separate vertical axes show the background-subtracted number of events for two different choices of the integrated luminosity: $138~\fb$ (Run~2) and $3~\ab$ (HL-LHC). The light and dark shaded gray 
areas show the statistical uncertainty on the SM background corresponding to the two luminosity assumptions. 
\begin{table}
\centering
\begin{tabular}{cc |c | c}
\multicolumn{2}{c}{$a$} & $A$ \\
$\ct/f_a \,  [\text{TeV}^{-1}]$ & $\cG/f_a \,  [\text{TeV}^{-1}]$ & $g_{A\ttbar}$ & $(\sigma^\text{tot}-\sigma^\text{SM})$ [pb] \\
\hline
\hline
$3.0$ & $+0.015$ & $0.95$ & $+6.7$ \\
$3.0$ & $-0.015$ & $0.43$ & $-2.7$ \\
$1.0$ & $+0.025$ & $0.75$ & $-1.7$ \\
$1.0$ & $-0.025$ & $0.87$ & $+2.0$ \\
\end{tabular}
\caption{BSM cross-section contributions in the four considered parameter benchmarks for the ALP and pseudoscalar Higgs boson. The coupling $\gAtt$ of the pseudoscalar Higgs boson is chosen such that its total cross-section contribution matches the one
of the ALP for each benchmark.}
\label{tab:benchmarks}
\end{table}

For the pseudoscalar Higgs boson $A$ (dashed) a characteristic peak--dip structure located around the particle mass
occurs in the $\mtt$ distribution for sufficiently large values of $\gAtt$ (upper left and lower right plot).
For small values of $\gAtt \lesssim 0.8$ 
(upper right and lower left plot), the depth of the dip, induced by the interference term, dominates over the height of the peak, which 
is mostly caused by
contributions from the resonance term, leading to a deficit of events throughout most of the considered $\mtt$ distribution.

Comparing the distributions of an ALP where $\cG \neq 0$
with those of a pseudoscalar Higgs boson,
differences in the $\mtt$ distributions become visible.
In particular, if $\cG$ and $\ct$ have opposite sign, instead of the peak--dip structure
a dip--peak structure may occur (upper right and lower right plots
in Fig.~\ref{fig:distributions}).
As expected,
the differences between 
the distributions for an ALP and a pseudoscalar Higgs boson
become more pronounced with increasing $|\cG|$
(lower left and lower right plots).
For the upper left and upper right plots, the $a$ and $A$
distributions are qualitatively similar, featuring a peak--dip and dip--dominated structure, respectively. However, the position and depth of the dip as well as the high-mass tail of the distribution are different for the case where $\ct$ and $\cG$ have opposite signs (upper right plot).
In the lower left plot, the ALP features a peak--dip structure, 
while the corresponding distribution for a
pseudoscalar Higgs boson is dominated by a dip only.
The most prominent 
difference in the distributions for an ALP and a pseudoscalar Higgs boson
can be observed in the lower right plot.
In this case, the $m_{\ttbar}$ distribution of the pseudoscalar Higgs boson features a peak--dip structure, while for the ALP a dip--peak structure occurs. 
In comparison to the indicated statistical uncertainties on the SM background for the two luminosity assumptions one can see that there is a certain sensitivity for discriminating between the hypothetical observation of an ALP and of a pseudoscalar Higgs boson already with the projection of the CMS analysis to the full Run~2 luminosity, and good prospects for all displayed scenarios for the case of the HL-LHC. These findings will be further quantified in the following.

In Tab.~\ref{tab:signif_ALP}, we present the expected significances for the observation of an ALP or a pseudoscalar Higgs boson for three different integrated luminosities corresponding to Run~2, Run~2+3 and the HL-LHC.
Analogously to Section~\ref{sec:syst_uncert}, we present the significances for three different assumptions on the uncertainties:
i) including statistical and all systematic uncertainties,
ii) including statistical and systematic uncertainties but excluding the top-quark mass uncertainty, 
iii) considering statistical uncertainties only. 
In addition, as discussed in
\refse{sec:syst_uncert},
we scale all systematic uncertainties by a factor of 0.5 for our projection to the HL-LHC in order to account for future improvements in prediction and analysis techniques.
We find that the benchmark scenario with $\ct/f_a = 3 \tevinv, 
\cG/f_a=0.015 \tevinv$ can be distinguished from the SM
expectation with a significance much above
$5\,\sigma$ at the HL-LHC for the case where all systematic and statistical uncertainties are taken into account. 
For all the other displayed benchmark scenarios an expected sensitivity at the HL-LHC above the level of $5\,\sigma$ can be achieved if the uncertainty arising from the top-quark mass in the analysis can be significantly reduced compared to our simple estimate (see the discussion above). As indicated in the ``no $m_t$ column'', for some of the displayed benchmark scenarios this level of significance could be reached in this case already with the integrated luminosity from Run~2 and Run~3.

Turning to
the question of how well an ALP and a pseudoscalar Higgs boson can be distinguished from one another in the considered benchmark scenarios,
the comparison of the 
predicted distributions for an ALP and a pseudoscalar Higgs boson
in Fig.~\ref{fig:distributions} with
the statistical uncertainty (gray bands) shows that for
the data that has been recorded at Run~2 
the deviation between an ALP and a pseudoscalar Higgs boson is largest in comparison to the statistical uncertainty for the benchmark scenario with
$\cG/f_a = -0.015\tevinv$, $\ct/f_a = 3.0\tevinv$ (lower right plot of Fig.~\ref{fig:distributions}). 
In this case the peak--dip structure caused by the pseudoscalar Higgs boson would be expected to be well separable from the background, while the ALP produced with the same total cross section would give rise to a dip--peak structure that is significantly less pronounced.
At the HL-LHC, all considered ALP benchmarks will be distinguishable from their pseudoscalar Higgs boson counterparts. 
More quantitatively, the significances for this comparison are given in Tab.~\ref{tab:signif_ALP_vs_Higgs}.
After LHC Runs 2(+3), only the benchmark scenario with $\ct/f_a = 1\tevinv$ and $\cG/f_a = -0.025 \tevinv$ has the potential to be distinguished from the case of a pseudoscalar Higgs boson with the same total cross section with (close to) $5\,\sigma$ significance.
For all four considered benchmark scenarios, a $5\,\sigma$ distinction of an ALP from a pseudoscalar Higgs boson with $\cG = 0$ will be possible at the HL-LHC, based on the result taking into account all systematic and statistical uncertainties. We note that in this case, as discussed above, the ALP signal itself may not be detectable with $5\,\sigma$ significance.

In case a new particle is detected at the LHC, the sensitivity for distinguishing between an ALP and a pseudoscalar Higgs boson would have an 
important impact on
the future collider programme
at the high-energy frontier. If one
can show that only
an ALP with $\cG \neq 0$ is in agreement
with the experimental data, 
this could imply the existence of additional
heavy BSM particles 
that are
responsible for the additional contributions
to the ALP--gluon contact interaction
in the ALP EFT. 
The size of $\cG / f_a$ which is consistent
with the data could then be used to
gain information on whether these BSM particles could potentially be
in reach of the LHC or other future colliders that are currently discussed.

\begin{table}
\centering
\begin{tabular}{cc |c||c|c|c|c}
\multicolumn{2}{c}{$a$} & $A$ &  & \multicolumn{3}{c}{Significance ($a$/$A$ vs. SM)} \\
$\ct/f_a \,  [\text{TeV}^{-1}]$ & $\cG/f_a \,  [\text{TeV}^{-1}]$ &  $\gAtt$ & Luminosity  & all syst. & no $m_t$ & stats only \\
\hline
\hline
\multirowcell{3}{$ 3.0$} & \multirowcell{3}{$+0.015$} 
& \multirowcell{3}{$0.95$}& Run 2 & $3.9$/$3.3$ & $> 10$/$8.9$ & $> 10$/$> 10$ \\
& & &  Run 2+3 & $5.2$/$4.3$ & $> 10$/$> 10$ & $> 10$/$> 10$ \\
& &  &  HL-LHC & $> 10$/$> 10$ & $> 10$/$> 10$ & $> 10$/$> 10$ \\
\hline
\multirowcell{3}{$ 3.0$} & \multirowcell{3}{$-0.015$} 
 & \multirowcell{3}{$0.43$}& Run 2 & $2.1$/$1.2$ & $2.2$/$1.5$ & $4.4$/$2.9$ \\
& & & Run 2+3 & $3.0$/$1.5$ & $3.0$/$2.0$ & $6.5$/$4.3$ \\
& & & HL-LHC & $8.7$/$4.2$ & $8.8$/$5.7$ & $> 10$/$> 10$ \\
\hline
\multirowcell{3}{$ 1.0$} & \multirowcell{3}{$+0.025$} 
 & \multirowcell{3}{$0.75$}& Run 2 & $1.1$/$2.4$ & $2.6$/$4.7$ & $4.0$/$6.3$ \\
& & & Run 2+3 & $1.4$/$3.1$ & $3.2$/$6.0$ & $5.9$/$9.4$ \\
& & & HL-LHC & $3.9$/$8.4$ & $8.2$/$> 10$ & $> 10$/$> 10$ \\
\hline
\multirowcell{3}{$ 1.0$} & \multirowcell{3}{$-0.025$} 
& \multirowcell{3}{$0.87$}& Run 2 & $0.7$/$2.8$ & $1.7$/$6.9$ & $2.8$/$9.8$ \\
& & & Run 2+3 & $0.9$/$3.6$ & $2.2$/$8.6$ & $4.1$/$> 10$ \\
& & & HL-LHC & $2.3$/$9.9$ & $5.5$/$> 10$ & $> 10$/$> 10$ \\
\end{tabular}
\caption{Significances for detecting an ALP or a pseudoscalar Higgs boson with a mass of 400~GeV and a width of 2.5\% for the benchmark scenarios considered in \reffi{fig:distributions}. Three different 
treatments of the uncertainties as defined in \refse{sec:syst_uncert} are shown. For the HL-LHC projection, all systematic uncertainties are scaled by a factor of 0.5. The ``/" separates the significances of the ALP from those of the pseudoscalar Higgs boson.
}
\label{tab:signif_ALP}
\end{table}

\begin{table}
\centering
\begin{tabular}{cc |c||c|c|c|c}
\multicolumn{2}{c}{$a$} & $A$ &  & \multicolumn{3}{c}{Significance ($a$ vs. $A$)} \\
$\ct/f_a \,  [\text{TeV}^{-1}]$ & $\cG/f_a \,  [\text{TeV}^{-1}]$ &  $\gAtt$ & Luminosity  & all syst. & no $m_t$ & stats only \\
\hline
\hline
\multirowcell{3}{$ 3.0$} & \multirowcell{3}{$+0.015$} 
& \multirowcell{3}{$0.95$}& Run 2 & $1.3$ & $1.9$ & $3.3$ \\
& & & Run 2+3 & $1.8$ & $2.3$ & $4.9$ \\
& & & HL-LHC & $5.3$ & $5.7$ & $> 10$ \\
\hline
\multirowcell{3}{$ 3.0$} & \multirowcell{3}{$-0.015$} 
 & \multirowcell{3}{$0.43$}& Run 2 & $1.2$ & $1.9$ & $3.3$ \\
& & & Run 2+3 & $1.7$ & $2.4$ & $4.9$ \\
& & & HL-LHC & $5.0$ & $6.0$ & $> 10$ \\
\hline
\multirowcell{3}{$ 1.0$} & \multirowcell{3}{$+0.025$} 
 & \multirowcell{3}{$0.75$}& Run 2 & $1.5$ & $2.3$ & $2.7$ \\
& & & Run 2+3 & $2.0$ & $3.1$ & $3.9$ \\
& & & HL-LHC & $5.8$ & $8.8$ & $> 10$ \\
\hline
\multirowcell{3}{$ 1.0$} & \multirowcell{3}{$-0.025$} 
& \multirowcell{3}{$0.87$}& Run 2 & $3.7$ & $9.0$ & $> 10$ \\
& & & Run 2+3 & $4.6$ & $> 10$ & $> 10$ \\
& & & HL-LHC & $> 10$ & $> 10$ & $> 10$ \\
\end{tabular}
\caption{Significances for the discrimination of an ALP and a pseudoscalar Higgs boson for the benchmark scenarios considered in \reffi{fig:distributions}. The uncertainties are treated as in Tab.~\ref{tab:signif_ALP}.}
\label{tab:signif_ALP_vs_Higgs}
\end{table}

\subsection{Projected ALP limits}

As seen in Fig.~\ref{fig:distributions},
the LHC results of Run~2 and Run~3
and especially of
the future high-luminosity phase
are expected to yield significant improvements 
of the sensitivity to
the ALP couplings $\cG$ and $\ct$. To quantify this, we 
derive estimates for the projected limits
on the ALP couplings from the investigated
$a \to \ttbar$ searches. 

To this end, we use the same uncertainty setup as presented in \refse{sec:syst_uncert} for computing significances. 
In a first step, we do not incorporate the systematic uncertainty arising from the top-quark mass.
We assume that the observed data is equal to the SM expectation, i.e.\ that no deviation from the SM is found, and scan over values of the ALP couplings $\cG/f_a$ and $\ct/f_a$, each time performing a $\mathrm{CL_s}$ test~\cite{Junk:1999kv,Read:2002al} using the test statistic given in \refeq{eq:teststat}. We assume the test statistic to be $\chi^2$-distributed with two degrees of freedom, and reject a set of values for the ALP couplings at 95\% confidence
level~(CL) if the $\mathrm{CL_s}$ value for these couplings falls below a threshold of 0.05.

\begin{figure}
\centering
\includegraphics[width=0.49\textwidth]{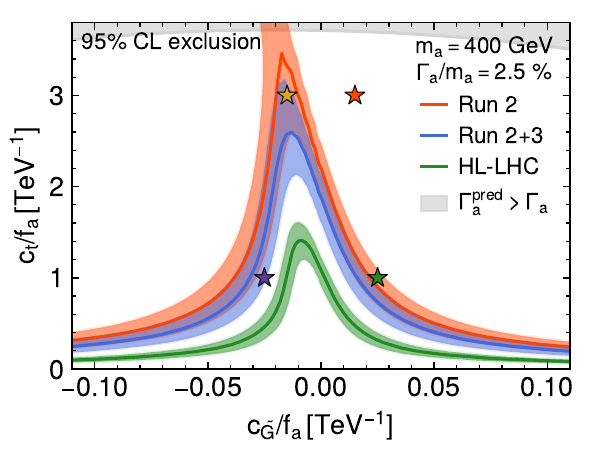}
\hfill
\includegraphics[width=0.49\textwidth]{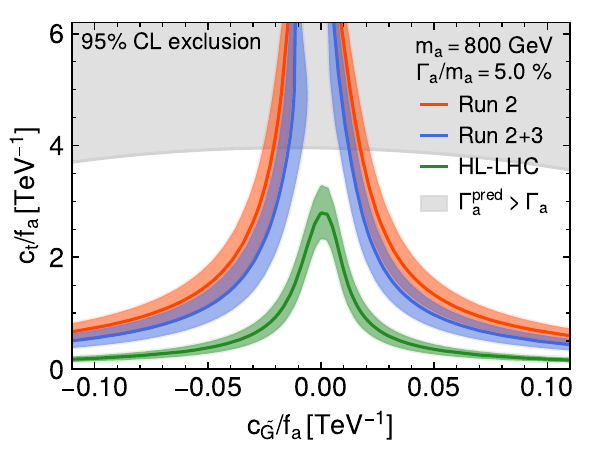}
\caption{Projected expected limits on the ALP couplings $\cG/f_a$ and $\ct/f_a$ as obtained by the maximum likelihood fit for three different integrated luminosities, corresponding to LHC Run 2 (red), 
Run 2+3 (blue) and HL-LHC (green), and for 
$m_a = 400 \gev$, $\Gamma_a/m_a = 2.5\%$ (left) and 
$800 \gev$, $\Gamma_a/m_a = 5.0\%$ (right). The shaded bands show the variations of the expected limit by one standard deviation. For all cases, all systematic uncertainties are included (scaled by a factor of 0.5 for the HL-LHC) except for the top-quark mass uncertainty. 
The region in which the assumed ALP width is lower than the total predicted width, taking into account the $a \to \ttbar$, $a \to gg$, $a \to Zh$ and $a \to \gamma\gamma$ decays,
is shown as the gray shaded area.
In the left plot, the benchmark points from \reffi{fig:distributions}, with the same color coding, are shown as stars. 
}
\label{fig:projectedlimits}
\end{figure}

The projected limits resulting from this procedure are shown in Fig.~\ref{fig:projectedlimits} for 
$m_a = 400 \gev$, $\Gamma_a/m_a = 2.5\%$ (left plot) 
and $\ma = 800 \gev$, $\Gamma_a/m_a = 5\%$ (right plot). In both cases, we include all systematic uncertainties except for the
top-quark mass uncertainty, as discussed in \refse{sec:syst_uncert}.
For Run 2, we find a limit of $c_t/f_a \leq 3.5 \tevinv$ 
for $m_a = 400 \gev$ in the least sensitive case for $\cG$ (corresponding to values of $\cG/f_a = -0.02 \tevinv$), while the limit 
is improved to $\ct/f_a \leq 0.34 \tevinv$ for $|\cG|/f_a = 0.1 \tevinv$.
For $m_a = 800 \gev$, we find a limit of $\ct/f_a \leq 0.7 \tevinv$ for $|\cG|/f_a = 0.1 \tevinv$, while no limit for the assumed total width can be set for $\cG = 0$.
The fact that the lowest sensitivity on $\ct$ is reached for a non-zero value of $\cG$ for $m_a = 400 \gev$ results from a destructive signal--signal interference between the two possible production diagrams, which suppresses the signal cross section for small negative values of $\cG$.
The four points indicated by stars in the left plot correspond to the four benchmark scenarios considered in
\reffi{fig:distributions}. The red benchmark point and possibly also the green one
can be probed already with the data from Run~2. 
The yellow benchmark point should become accessible with the integrated luminosity after Run~3 of the LHC, while the purple one becomes only accessible at the HL-LHC.

It should be noted that for $m_a = 800 \gev$  and $\Gamma_a/m_a = 5\%$, similar to \reffi{fig:cmslimits}, the expected exclusion limits on $\ct/f_a$ for Run 2 
and Run 3 within the interval
$|\cG|/f_a \leq 0.02$
lie in the region where the predicted total ALP decay width exceeds the width assumed in the analysis.\footnote{If we instead choose the total width as $\Gamma_a = \Gamma_{\ttbar}^{\mathrm{pred}} + \Gamma_{g g}^{\mathrm{pred}} + \Gamma_{\gamma \gamma}^{\mathrm{pred}} + \Gamma_{Z h}^{\mathrm{pred}}$, we expect our limits to be weaker in the gray-shaded region and to be stronger otherwise.} 
For the 
integrated luminosities expected at the HL-LHC
limits for both ALP masses can be set throughout the whole parameter 
space without a conflict with the assumed value for the total width.
For the HL-LHC we
find a significantly improved limit of $\ct/f_a \leq 1.4 \, (2.8) \tevinv$ 
for $m_a = 400 \, (800) \gev$.
For $|\cG|/f_a = 0.1 \tevinv$, we find a limit of $\ct/f_a \leq 0.11 \, (0.19) \tevinv$. 

\begin{figure}
\centering
\includegraphics[width=0.49\textwidth]{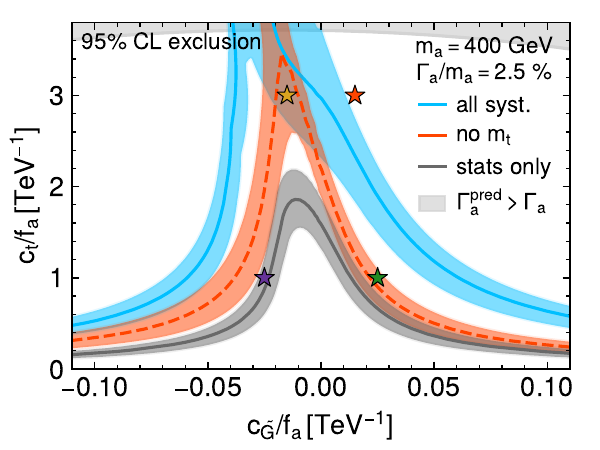}
\hfill
\includegraphics[width=0.49\textwidth]{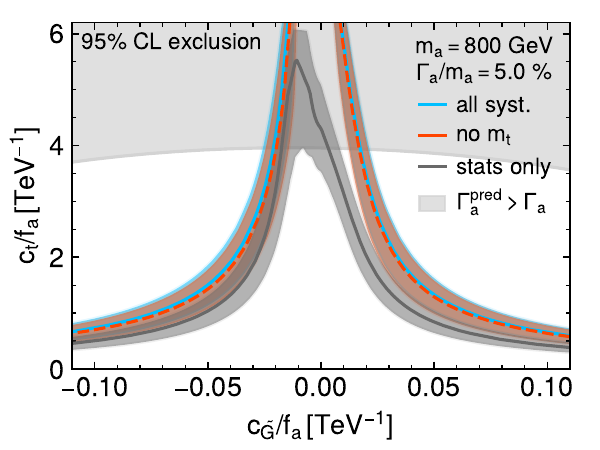}
\caption{Projected expected limits on the ALP couplings $\cG/f_a$ and $\ct/f_a$ that are obtained as in
\reffi{fig:projectedlimits}, for $m_a = 400 \gev$, $\Gamma_a/m_a = 2.5\%$ (left plot) 
and $\ma = 800 \gev$, $\Gamma_a/m_a = 5\%$ (right plot) and for three different treatments of the uncertainties as defined in \refse{sec:syst_uncert}: with all systematic uncertainties including the top-quark mass uncertainty (blue), excluding the top-quark mass uncertainty but keeping other uncertainties (red), 
and considering statistical uncertainties only (gray). All limits are shown for an integrated luminosity of $138 \, \fb$, corresponding to Run 2.
}
\label{fig:limits_systs}
\end{figure}

We now analyze the impact of the treatment of the uncertainties, in particular the effect of
taking into account the 
systematic uncertainty arising from the top-quark mass.
To this end, 
in \reffi{fig:limits_systs} we show the same limits for an integrated luminosity corresponding to Run 2 for the three
different treatments of the uncertainties
as in Tabs.~\ref{tab:signif_ALP}--\ref{tab:signif_ALP_vs_Higgs}. For $m_a = 400 \gev$ (left plot), i.e.\ close to the $\ttbar$ production threshold, it can be seen that including the top-quark mass uncertainty significantly weakens the projected limits. As discussed in \refse{sec:syst_uncert}, this uncertainty is likely overestimated in our analysis, and we expect the limits that would be found in an experimental analysis to lie between the cases including or excluding the top-quark mass uncertainty (blue and red lines).

For $m_a = 800 \gev$ (right plot), on the other hand, it can be seen that the effect of the top-quark mass uncertainty is very small.
For this mass, the signal does not manifest itself as a peak--dip structure close to the $\ttbar$ production threshold 
and thus
has a very different
shape compared to modifications of the
$m_{t \bar t}$ distribution caused by
a variation of the mass of the top quark.

\subsection{Comparison with other
experimental limits}
\label{sec:bounds}

\begin{figure}[t]
    \centering
    \includegraphics[width=.8\textwidth]{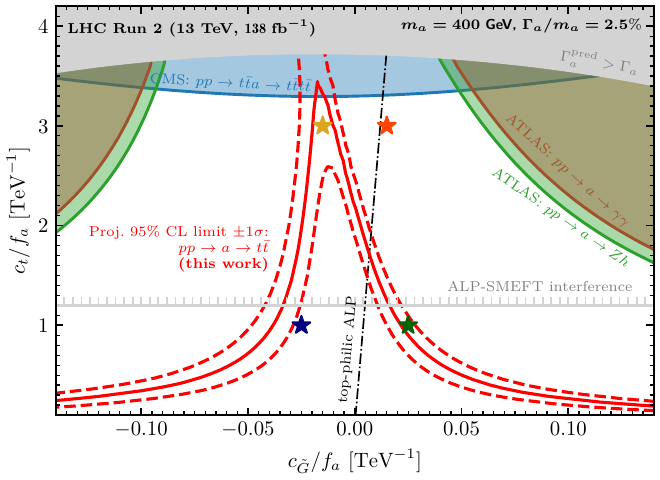}
    \caption{
    Limits on the ALP Wilson coefficients $\ct/f_a$ and $\cG/f_a$ from various sources:
    The red line shows our result for the
    projected expected 95\% CL upper
    limit on the
    ALP couplings $\cG/f_a$ and $\ct/f_a$ from
    ALP searches in the invariant $t \bar t$
    mass distribution and angular
    correlations for an ALP with a mass of
    $400\gev$ and relative
    width of $2.5\%$ 
    for the case where all statistical and systematic uncertainties 
    are included except for the top-quark mass uncertainty
    (see Fig.~\protect\ref{fig:projectedlimits}). The red dashed
    lines indicate the $\pm 1\sigma$
    uncertainty band.
    The brown shaded areas are excluded
    at 95\% CL from LHC searches for narrow
    di-photon resonances performed by
    ATLAS~\cite{ATLAS:2021uiz},
    the green shaded areas are excluded
    from LHC searches for pseudoscalar
    resonances decaying into a $Z$ bosons
    and the Higgs boson at 125~GeV
    performed by ATLAS~\cite{ATLAS:2022enb},
    and the blue shaded area is excluded
    from the CMS cross section measurement
    for the production of four top
    quarks~\cite{CMS:2023ftu},
    where both analyses
    are based on the full Run~2 dataset.
    Values of $c_t / f_a$ above the gray
    hatched line are excluded by ALP--SMEFT
    interference 
    effects~\cite{Biekotter:2023mpd}
    (see text for details).
    Coupling values for which the
    predicted sum of the partial widths for 
    $a \to \ttbar, gg, \gamma\gamma$
    is larger
    than the assumed total width 
    of the ALP are indicated with the gray shaded area. The black dot-dashed line represents the couplings corresponding to a top-philic ALP.
    }
    \label{fig:comprun2}
\end{figure}

In this section we compare the projected limits
on the ALP couplings obtained in the previous
section to other
current experimental limits on these parameters.
Focusing on an ALP at $m_a = 400\gev$ and a
relative width of $\Gamma_a / m_a = 2.5\%$,
we show in \reffi{fig:comprun2} our projected
95\% CL limits on the coupling coefficients
$\cG/f_a$ and $\ct/f_a$ based on the $pp \to a \to t \bar t$ process for an integrated luminosity of
$138~\mathrm{fb}^{-1}$ collected during Run~2 
(as shown by the red curve in \reffi{fig:projectedlimits}).
For the specific scenario of the top-philic ALP (see \refeq{eq:top-philic}), indicated by the black dot-dashed line, we find from the intersection with the red curve a projected limit of $\ct/f_a < 1.7 \tevinv$, which is complementary to the limit set at lower masses in \citere{Blasi:2023hvb}.
Our limits are compared with the current limits resulting from LHC searches for di-photon
resonances~\cite{ATLAS:2021uiz} (brown
shaded area), searches for pseudoscalar
resonances decaying into a $Z$ boson and
a 125~GeV Higgs boson (green shaded area)~\cite{ATLAS:2022enb},
and from the measurement of the total
cross section
for the production of four top
quarks~\cite{CMS:2023ftu} (blue shaded area), all
based also on $138~\mathrm{fb}^{-1}$. Moreover,
we depict the indirect limit on $\ct/f_a$, which is
approximately independent of $\cG/f_a$ for $m_a = 400\gev$,
resulting from a global analysis of
ALP--SMEFT interference
effects~\cite{Biekotter:2023mpd} (hatched gray line).
In the following we briefly summarize
how the existing
exclusion regions displayed in \reffi{fig:comprun2}
were obtained and discuss the assumptions on
which they are based.

\medskip

\noindent\textbf{Direct searches for resonant
$gg \to a \to \gamma\gamma$ production}

\noindent

As explained above, the ALP coupling to top quarks induces at the one-loop level a coupling of the ALP to two photons, see 
\refeq{eq:gayyeff}.
Searches for the ALP as a narrow di-photon
resonance profit from a relatively small background, 
while interference effects between
ALP production and the SM background are negligible.
Existing LHC searches for high-mass di-photon 
resonances can therefore be used as a different way for probing
ALPs as considered here.
However, the di-photon branching ratio is strongly
suppressed for ALP masses above the $\ttbar$ threshold.
For an ALP mass of $m_a = 400\gev$ we can apply
the 95\% 
CL
cross-section
limits resulting from
the ATLAS high-mass di-photon
resonance search~\cite{ATLAS:2021uiz}
as implemented in
HiggsTools~\cite{Bechtle:2008jh,
Bechtle:2011sb,
Bechtle:2013wla,Bechtle:2015pma,Bechtle:2020pkv,
Bahl:2022igd}.\footnote{The 
corresponding CMS search~\cite{CMS:2018dqv}
sets limits for narrow
di-photon resonances with masses above
500~GeV, and 
therefore does not apply to the considered scenario.}
We obtain
the theoretical prediction for the
resonant cross section of the ALP in the same way as described in \refse{sec:vegas}, i.e.~we
compute
the resonant gluon-fusion production cross-section at~LO
with \textsc{MadGraph 5} and apply a
$K$-factor
$K_{\mathrm{res}}$
to account for higher-order
QCD effects.
As described in \refse{sec:widths}, the prediction
for the decay $a \to \gamma \gamma$ 
is based on the loop-induced coupling involving the top-quark loop.

We stress again in this context
that we treat
the total width of the ALP as a free parameter in our analysis.
Accordingly, for the coupling regions below the gray area
in \reffi{fig:comprun2}, where the predicted sum of the partial
widths for the $t \bar t$, $gg$ and $\gamma \gamma$ decays
is smaller than the assumed total width,
the ALP would necessarily feature
additional couplings to other SM or BSM particles.
If those additional couplings of the ALP involve charged particles, the partial width for $a \to \gamma\gamma$ would receive additional loop-induced contributions besides
the contribution from the top-quark loop.
As a consequence,
the exclusion regions resulting from the
di-photon searches shown in
\reffi{fig:comprun2} can be significantly modified via the impact of
additional ALP couplings on the di-photon branching ratio.

\medskip

\noindent\textbf{Direct searches for resonant
$gg \to a \to Z h$ production}

\noindent

A pseudoscalar can also decay into
a $Z$ boson and a CP-even scalar.
For the ALP considered here, and assuming
that the Higgs boson~$h$ at 125~GeV is a CP-even
scalar as predicted by the SM, the
decay $a \to Zh$ is generated at the one-loop
level resulting from the interaction with
the top quark.
The most sensitive experimental
searches for this signature
utilize the decay of $h$ into
a pair of bottom quarks. These
have been performed at 13~TeV
by both the ATLAS and the CMS collaborations
using first-year Run~2
data~\cite{ATLAS:2017xel,CMS:2019qcx} and more recently
by ATLAS using the full Run~2
data set~\cite{ATLAS:2022enb}, providing
the currently strongest limits
on the $pp \to a \to Z h$ 
process.\footnote{Recently,
CMS has reported the results from
$a \to Zh$ searches with $h$ decaying into
a pair of $\tau$-leptons
taking into account the full Run~2
data set~\cite{CMS-PAS-HIG-22-004}.
The resulting cross-section limit at about
400~GeV is $\sigma(gg \to a \to Zh) \approx
0.4~\mathrm{pb}$, which is very similar to the
corresponding limit obtained
by CMS using the $h \to b \bar b$
decay mode taking into account
first-year Run~2 data only~\cite{CMS:2024phk},
and about a factor of four weaker than
the ATLAS limit including the full
Run~2 data set~\cite{ATLAS:2022enb}.}
To apply these limits to the ALP we implemented them
into \texttt{HiggsTools}. We find that the
searches for $a \to Zh$
give rise to very similar exclusion regions
in \reffi{fig:comprun2} as the ones from the
$a \to \gamma\gamma$ channel.
Here it should be kept in mind, as discussed
above for the di-photon decay, that the predicted
branching ratio for the
loop-induced decay $a \to Zh$ is very sensitive
to the assumptions on possible additional couplings
of the ALP (which we assume to vanish
for the displayed exclusion regions).

\medskip
\noindent\textbf{Direct sensitivity to $a \to t \bar t$
in the production of four top quarks}

At $m_a = 400\gev$ and assuming
$\Gamma_a / m_a = 2.5\%$, the ALP investigated
here mainly decays into top-quark pairs. 
For the case where
the ALP is produced in association with two top quarks,
this decay mode contributes
to the production of four
top quarks at the LHC.
This final state was recently measured
for the first time
using the Run~2 dataset collected
at~13~TeV by both the
CMS~\cite{CMS:2023ftu,CMS:2023zdh}
and the
ATLAS~\cite{ATLAS:2021kqb,ATLAS:2023ajo}
collaborations.
These measurements can be used to set
upper limits
on the process
$pp \to t \bar t a \to t \bar t t \bar t$.
The exclusion region shown in
\reffi{fig:comprun2} is based
on the CMS measurement in the same-sign di-lepton
plus multilepton channel~\cite{CMS:2023ftu},
which has the
smallest statistical and
systematic uncertainties among the
currently existing measurements. 
CMS found
a cross section of $\sigma_{t \bar t t \bar t} =
17.7^{+ 4.4}_{-4.0}~\mathrm{fb}$, which agrees
with the SM prediction
$\sigma_{t \bar t t \bar t}^{\rm SM} =
13.4~\mathrm{fb}$~\cite{vanBeekveld:2022hty}
at about the level of $1\,\sigma$.
We use as an upper limit
on the cross section the upper value of the
$2\,\sigma$ uncertainty band, which corresponds
approximately to a 95\% CL limit.
As a simple estimate of
the theoretical prediction for the ALP model,
the cross section for
$pp \to t \bar t a \to t \bar t t \bar t$ 
is added to the cross section that is predicted 
within the SM. This is conservative in the sense that 
taking into account a lower acceptance 
for events arising from the resonant process of ALP 
production than for the non-resonant 
SM-like contribution is expected to 
reduce the impact of this 
constraint on the displayed parameter space. 
Since 
for the final state with four top quarks the
interference effects 
between the resonant ALP contribution and the non-resonant SM background
are much less important compared to the
case of $t \bar t$ production, 
they have been neglected here.
The cross section for the production of the ALP
in association with two top quarks was obtained
with the help of \texttt{HiggsTools} as a function
of $c_t$, and the decay width for $a \to t \bar t$
was computed according to \refeq{eq:widtt}
as in our analysis for the $\ttbar$ final state.
The resulting exclusion limit only mildly
depends on $\cG$ since $\cG$ enters
this process
only in the branching ratio for the $a \to \ttbar$
decay, which is the dominant decay mode
for the considered coupling values.
We find that the exclusion limit from the measurement of 
the production of four top quarks is significantly weaker than
our projected limit from the $t \bar t$ searches except for the region of small negative values of 
$\cG/f_a$ where the latter limit is weakest.
The limit on $c_t / f_a$ that we obtain
for $\cG = 0$ can be furthermore compared to limits
that CMS has obtained
in a search for new spin-0 particles in
final states with three or four top quarks
published in
\citere{CMS:2019rvj}.
CMS found in the 2HDM interpretation
a lower limit of $\tan\beta = 1.2$ at
a mass of 400~GeV (see upper right plot
of Fig.~8 therein), which in terms
of the ALP Wilson coefficients corresponds to
a limit of $\ct / f_a = 3.4 \tevinv$. This is in
good agreement with the limit that we have
obtained using the total cross section
measurement of \citere{CMS:2023ftu}.

\medskip

\noindent\textbf{Indirect effects from ALP--SMEFT interference}

ALP couplings can also be constrained indirectly through their impact on observables described in the SMEFT framework. The RG
evolution induces non-zero SMEFT coefficients 
at scales probed at LEP or the LHC
even if the ALP couplings are the only BSM contributions present at the UV scale~\cite{Galda:2021hbr}.
This effect has been used to constrain the Wilson coefficients of the ALP effective Lagrangian by reinterpreting SMEFT constraints from LHC Higgs and top data as well as electroweak precision observables. 
The resulting bounds on ALP couplings to gluons and top quarks are~\cite{Biekotter:2023mpd}\footnote{It should be noted that these bounds 
apply to
the couplings at the high scale, $c_x(\Lambda)$, rather than to the ones
at the ALP mass $m_a$.} 
\begin{align}
\label{eq:indirectlimits}
|c_t|/f_a < 1.2 /\text{TeV}, \, \qquad  |c_{\tilde G}|/f_a < 0.74/\text{TeV}  \, , \quad 
        & {\rm for } \; m_a = 400 \, \text{GeV} 
        \nonumber \\
|c_t|/f_a < 1.3/\text{TeV} , \,  \qquad |c_{\tilde G}|/f_a < 0.83 /\text{TeV} \, , \quad  
        & {\rm for } \; m_a = 800 \, \text{GeV}  \, .
\end{align}
The limit on $\ct$ is dominated by 
its contribution to
the SMEFT Wilson coefficient $C_{HD}$ corresponding to the SMEFT operator 
\begin{align}
  \mathcal{O}_{HD} =  (H^\dagger D_\mu H)^* (H^\dagger D^\mu H)  \, ,
\end{align}
which is tightly constrained from electroweak precision observables, most notably from the $W$-boson mass.\footnote{These limits
are obtained using the experimental 
average value of $M_W = 80.379 \pm 0.012$~GeV~\cite{Zyla:2020zbs}.
This value does not include the recent CDF measurement~\cite{CDF:2022hxs}
which is in significant tension with the SM.}
For $\cG$ the reinterpretation of Higgs limits on $C_{uG}$, $C_{HG}$, corresponding to the dimension-six SMEFT operators 
\begin{align}
    \mathcal{O}_{HG} = \left(\phi^\dagger \phi  \right) \, G^a_{\mu \nu} G^{a\mu \nu} \, , 
    \quad 
    \mathcal{O}_{uG} = \left( \bar{Q}_L \sigma_{\mu \nu} u_R \right)  
    T^a \tilde{\phi} \, G^{a \mu \nu} \, ,
\end{align}
dominates the bounds. 
While these bounds are independent of, for instance, the ALP decay width or its branching ratios, they assume that all SMEFT Wilson coefficients are exactly zero at the high scale $\Lambda = 4 \pi f_a$. 
However, the presence of additional non-zero SMEFT Wilson coefficients at $\Lambda$ can influence the limits on the ALP coefficients $c_t$ and $c_{\Tilde{G}}$ in either direction. 

\medskip

One can see in \reffi{fig:comprun2} that
the expected limits from an investigation of the invariant $t \bar t$ mass
distribution (red line)
are substantially stronger than current limits
from LHC searches for narrow
resonances decaying into two photons
(brown shaded area) or into
a $Z$ boson and a 125~GeV Higgs boson
(green shaded area),
and from the cross
section measurement for the production
of four top quarks (blue shaded area). 
Only in the region where the limits from the $m_\ttbar$ distribution become weak, i.e.\ for small values of $\cG$, the limits from the searches for
four top quarks become comparable.
The projected limits obtained in our analysis are
the only limits from direct searches at
the LHC that are comparable or stronger
than
the indirect limits on $\ct/f_a$ from the
ALP--SMEFT interference effects
(gray hatched line).
We find limits on $\ct / f_a$ that are
up to an order
of magnitude stronger 
than the indirect
bound on $\ct / f_a$ for values of $|\cG| / f_a \gtrsim 0.05 \tevinv$,
whereas for smaller values of $|\cG| / f_a$
the indirect limit from the ALP--SMEFT
interference is stronger.

It should be noted here that all limits from direct searches depend on the assumed total width of $\Gamma_a/m_a = 2.5\%$ via the branching ratio into the respective final states. If a lower width is assumed, including the case where $\Gamma_a$ is given only by the $\ttbar$, $gg$, $\gamma\gamma$, and $Zh$ decays, all direct searches are expected to give stronger limits, while the 
relative impact of the different decay channels will roughly stay the same.

\section{Summary and Conclusion}
\label{sec:conclusion}

Axion-like particles (ALPs) are singlets under the 
SM gauge groups
and appear in many well-motivated extensions of the SM
as the lightest degree of freedom due to
their nature as
pseudo-Nambu-Goldstone bosons of an approximate
axion shift-symmetry.
Therefore, ALPs are an attractive target for the 
LHC and the HL-LHC in the hunt for BSM physics. 
In this paper we studied the gluon-fusion production
of an ALP at the LHC with subsequent decay into
a pair of top quarks. This channel is
directly sensitive to
both the ALP--fermion and the ALP--gluon couplings
in the production, and to the ALP--fermion coupling
in the decay. Motivated by recent searches for
additional Higgs bosons in $t \bar t$ final states
by the ATLAS and CMS collaborations,
we have analyzed the current limits and future prospects
for probing the parameter space of the effective
ALP Lagrangian.

We have performed a recasting of the published CMS search for a CP-odd 
Higgs boson, extending it to the case of an ALP with a more general coupling structure.
In our simulation
the decay of the top quarks has been included,
thus going beyond previous
phenomenological studies
of ALP searches in $m_{t \bar t}$ distributions.
Since top quarks decay on timescales much
smaller than the one of the strong interaction,
angular variables of the decay products
can be used to gain sensitivity to spin
information of the
top quarks. This information can be utilized to enhance the sensitivity for 
discriminating a signal from the background and for characterizing the 
properties of a possible signal.  
Exploiting the information from 
the invariant-mass distribution
of the final-state top quarks,
$m_{t \bar t}$, and the spin
correlation variable, $\chel$,
we have investigated in particular how the
production of an ALP
with generic couplings to gluons and top quarks 
can be distinguished 
from the production of a pseudoscalar which couples to gluons exclusively 
via a top-quark loop.

In order to incorporate the effects of a 
finite detector resolution into our phenomenological analysis,
we have applied a Gaussian smearing with $\sigma = 15 \%$ 
on the $\mtt$ distribution. We determined the 
appropriate magnitude of the smearing
from a fit of the smeared generator predictions 
to both the SM $\ttbar$ background and the expected signal for 
a CP-odd Higgs-boson that was obtained in the CMS analysis based on a  
full detector simulation.
By comparing the distributions with and without imposing the cut 
$\chel > 0.6$ on the helicity variable we have demonstrated the high 
discrimination power of this variable. 

In our analysis we have investigated in detail different
sources of 
systematic uncertainties. In this context we have pointed out in particular 
the importance of the systematic uncertainty that is associated with the 
uncertainty on the mass of the top quark, which strongly affects 
the $\mtt$ distribution for the SM $\ttbar$ background in the low-mass 
region just above the $\ttbar$ threshold. In our phenomenological analysis 
we have used a Gaussian uncertainty of $\pm 1 \gev$ for the top-quark mass.
For the example of an expected signal of a CP-odd Higgs boson at 400~GeV we
have demonstrated that the variation of the top-quark mass in the SM 
background by $-1 \gev$ yields, after subtracting the SM background with 
$m_t = 172.5 \gev$, a pattern resembling the peak--dip structure that is 
expected for the signal. Since in our phenomenological analysis, which does 
not take into account variations in the acceptance arising from the top-quark mass dependence, the uncertainty associated with the top-quark mass is likely to be overestimated, we have presented our results with and without the uncertainty stemming from
the top-quark mass.
In order to compute expected significances and limits including the systematic uncertainties,
we have performed hypothesis tests based on a binned profile likelihood fit.

As a first step in our numerical analysis we have employed the 
results from the CMS search for a CP-odd Higgs boson using 
$35.9$~fb$^{-1}$ of data collected at $\sqrt{s} = 13$~TeV (with a similar expected sensitivity as the preliminary Run~2 ATLAS result) to derive 
limits on the effective ALP
Lagrangian in terms of the Wilson
coefficient $\ct$ for the case $\cG = 0$. For the example of 
an ALP in the mass range of 
$400\gev < m_{a} \lesssim 550\gev$ and with a relative width of 5\%,
coupling values of $|\ct| / f_a \gtrsim 4\tevinv$ are excluded. 
We have then investigated the expected significances for the observation of 
an ALP or a pseudoscalar Higgs boson for the integrated luminosities 
corresponding to Run~2 and Run~2$+$3 of the LHC as well as to the HL-LHC. 
For a benchmark scenario with $\ct/f_a = 3 \tevinv$ and 
$\cG/f_a=0.015 \tevinv$, taking into account all systematic 
and statistical uncertainties, we have shown that the discrimination from the SM expectation is possible with very high significance at the HL-LHC.
For all the other investigated benchmark scenarios 
for ALPs and pseudoscalar Higgs bosons we have found that
an expected sensitivity at the HL-LHC above the level of $5\,\sigma$ can be achieved if the uncertainty arising from the top-quark mass in the analysis can be significantly reduced compared to our simple estimate. In this case such a
level of significance could be reached already with the integrated luminosity from Run~2 and Run~3 for some of the investigated benchmark scenarios. 

As a further step we determined the significances for 
distinguishing a generic ALP from a pseudoscalar Higgs boson
that couples to gluons exclusively via a top-quark loop.
We have found that at the HL-LHC all considered ALP benchmarks will be distinguishable from their pseudoscalar Higgs boson counterparts for the case where the latter have the same mass and relative width as the considered ALP and where the couplings are such that the integrated cross sections for the two types of BSM particles are the same.
Already with the data from Run~2 and Run~3 of the LHC a significant sensitivity for distinguishing between an ALP and a pseudoscalar Higgs boson is achieved. We note that this kind of information can have important implications for the future collider programme at the high-energy frontier because of the different prospects for detecting additional BSM particles.

Turning from the prospects for discovering new particles to 
projected limits from ALP searches, we have determined 
projected limits on the ALP couplings
to fermions and gluons in terms of the
Wilson coefficients of the ALP--SM Lagrangian under 
the assumption that no deviations from the
SM expectation will be observed.
Including all systematic uncertainties except for the
top-quark mass uncertainty, we have found for
Run~2 a projected limit of $c_t/f_a \leq 3.5 \tevinv$ 
for $m_a = 400 \gev$ in the least sensitive case for $\cG$ (corresponding to values of $\cG/f_a = -0.02 \tevinv$), while the limit 
is improved to $\ct/f_a \leq 0.34 \tevinv$ for $|\cG|/f_a = 0.1 \tevinv$.
For $m_a = 800 \gev$ we have obtained a projected limit for Run~2 of $\ct/f_a \leq 0.7 \tevinv$ for $|\cG|/f_a = 0.1 \tevinv$. 
For the HL-LHC we
find a significantly improved projected limit of $\ct/f_a \leq 1.4 \, (2.8) \tevinv$ 
for $m_a = 400 \, (800) \gev$ and $\cG \approx 0$.
For $|\cG|/f_a = 0.1 \tevinv$, we have obtained a projected limit of $\ct/f_a \leq 0.11 \, (0.19) \tevinv$.
Regarding the impact of taking into account the systematic uncertainty arising from the top-quark mass, as expected we have found significant effects for the case of
$m_a = 400 \gev$, i.e.\ close to the $\ttbar$ production threshold, while for $m_a = 800 \gev$ this uncertainty has only a very small effect.

In order to assess the impact of our projected limits for Run~2 of the LHC, 
in a final step we have compared those limits from the $t \bar t$ searches 
(focusing on the case $m_a = 400 \gev$) with existing experimental
bounds from LHC searches for narrow di-photon resonances and for new resonances decaying into a Z boson
and a 125~GeV Higgs boson, and furthermore
from measurements of the production of four top quarks. We have also shown 
for comparison the constraints from global analyses of ALP--SMEFT 
interference effects, which arise from renormalization group running 
effects that induce a mixing between ALP EFT operators and SMEFT operators.
We showed that our projected limits from the $t \bar t$ searches are 
significantly stronger than the ones from the measurement of the production 
of four top quarks except for the region of small negative values of 
$\cG/f_a$, where our projected limit from the $t \bar t$ searches is 
weakest. For the largest values of $|\cG| / f_a$ considered in our analysis, 
our projected limits from the $t \bar t$ searches are up to
an order of magnitude stronger.
The current limits from searches for $a \to \gamma\gamma$
and $a \to Zh$
(for the considered case 
these decays are only generated via the top-quark
loop contribution)
have turned out to be always substantially weaker compared to our
projected limits from the $t \bar t$ searches.
In comparison to the indirect limits obtained from 
ALP--SMEFT interference effects, which rely on the
assumption that the ALP dimension five operators
are the only BSM contribution present at the UV~scale,
we have found that the projected direct limits from the $t \bar t$ searches 
at the LHC are the only ones that can give rise to limits
comparable or below the indirect limit on $\ct / f_a$
from ALP--SMEFT interference effects.
Overall, the limits on $\ct / f_a$
that can be obtained from
ALP searches in the $m_{t \bar t}$ distribution
are expected to be the strongest limits 
in the range $\cG / f_a < -0.04\tev^{-1}$ and
$\cG / f_a > 0.02\tev^{-1}$.

\bigskip

To conclude,
we derived limits on the ALP coupling to the top quark for ALPs with a mass above the $\ttbar$ threshold. First, we reinterpreted existing limits on pseudoscalar Higgs bosons for ALPs in the case where no contact interaction with gluons exists beyond the one induced by the top-quark loop.
As the main part of our analysis, we explored the sensitivities to distinguish ALPs from heavy pseudoscalar Higgs bosons. 
Within the considered benchmark scenarios, we find that a distinction can be possible already using the LHC Run~2 dataset, with substantially improved  prospects at the HL-LHC.
Assuming the absence of a signal, we derived expected limits dependent on the ALP--top and ALP--gluon couplings which are significantly stronger than existing direct limits from other searches. They also complement indirect limits derived from ALP--SMEFT interference. 
We encourage the experimental collaborations to adopt the  strategies outlined in our paper for future ALP searches at the LHC.



\section{Acknowledgements}
We thank Katharina Behr, Quentin Bonnefoy and Susanne Westhoff for useful discussions. 
A.B.\ is supported by  the  Cluster  of  Excellence  ``Precision  Physics,  Fundamental
Interactions, and Structure of Matter" (PRISMA$^+$ EXC 2118/1) funded by the German Research Foundation (DFG) within the German Excellence Strategy (Project ID 390831469). 
The work of T.B.~is supported by the German
Bundesministerium f\"ur Bildung und Forschung (BMBF, Federal
Ministry of Education and Research) -- project 05H21VKCCA.
The project that gave rise to these
results received the support of a
fellowship from the ``la Caixa''
Foundation (ID 100010434). The
fellowship code is  LCF/BQ/PI24/12040018.
The work of S.H.\ has received financial support from the
grant PID2019-110058GB-C21 funded by
MCIN/AEI/10.13039/501100011033 and by ``ERDF A way of making Europe", 
and in part by the grant IFT Centro de Excelencia Severo Ochoa CEX2020-001007-S
funded by MCIN/AEI/10.13039/501100011033. 
S.H.\ also acknowledges support from Grant PID2022-142545NB-C21 funded by
MCIN/AEI/10.13039/501100011033/ FEDER, UE.
A.G., L.J., C.S.\ and G.W.\ acknowledge support by the Deutsche
Forschungsgemeinschaft (DFG, German Research
Foundation) under Germany‘s Excellence
Strategy -- EXC 2121 ``Quantum Universe'' --
390833306.
This work has been partially funded
by the Deutsche Forschungsgemeinschaft 
(DFG, German Research Foundation) - 491245950. 

\appendix

\section{Further one-loop ALP interactions}

\label{app:moredecays}

In Section~\ref{sec:eff_couplings}, we analyzed the effective couplings of an ALP to gluons and photons. This appendix is dedicated to the calculation of further ALP couplings induced by the ALP--top coupling 
at the one-loop level
which lead to subdominant ALP decay 
channels.
We discuss contributions of ALP decays to $Z\gamma$ and charged leptons through
the mixing of the ALP with the longitudinal mode of the $Z$ boson.

\medskip

\noindent $\mathbf{a \to Z \gamma}$:

The decay $a \to Z \gamma$ arises
at the one-loop level via a top-quark loop. The
corresponding partial width divided by the
partial width for the di-photon decay is given by
\begin{equation}
  \frac{\Gamma(a \to Z\gamma)}{\Gamma(a \to \gamma\gamma)}
    = \frac{3 - 8 s_w^2}{8 c_w s_w}
    \cdot
    \frac{C_0(0,m_Z^2,m_a^2,m_t^2,m_t^2,m_t^2)}
      {C_0(0,0,m_a^2,m_t^2,m_t^2,m_t^2)}
    \, ,
\end{equation}
where $s_w$ and $c_w$ are the sine and the cosine
of the weak mixing angle, respectively,
$m_Z$ is the mass of the $Z$-boson,
and the one-loop three-point
function $C_0$ can be found
in \citeres{tHooft:1978jhc,Hahn:1998yk}.
For an ALP at 400~GeV one finds a ratio of
$\Gamma(a \to Z\gamma) / \Gamma(a \to \gamma\gamma) \approx 0.36$, i.e.~the branching ratio
for the di-photon decay is about a factor of three
larger.

The currently strongest limits
on this process were published
by the ATLAS collaboration using the full
Run~2 data set~\cite{ATLAS:2023wqy}.
In comparison to the searches for
di-photon resonances, the searches for
$a \to Z \gamma$ give rise to weaker 
experimental limits at the LHC.
Moreover, as explained above,
for the ALP at $m_a = 400\gev$
the branching ratio for $a \to Z \gamma$
is about a factor of three smaller than
the one for the di-photon decay.
As a result, for the ranges of the Wilson
coefficients displayed in \reffi{fig:comprun2}
no exclusion regions arise from
searches for $a \to Z\gamma$.

\medskip

\noindent $\mathbf{a \to \ell^+ \ell^-}$

Couplings of the ALP to charged leptons
$\ell^\pm$ are generated at the one-loop level due
to the mixing of the ALP with the $Z$ boson
and the neutral Goldstone boson resulting
from self-energy diagrams with a
top-quark loop~\cite{Bonilla:2021ufe}.
The effective ALP--lepton coupling is given by 
\begin{equation}
    g_{\ell}^\text{eff}  = \frac{\alpha}{2 \pi} c_t \left[ \frac{3 m_t^2}{2 s_w m_W^2} \left( \log \frac{\Lambda^2}{m_a^2} + 2 +i \pi \right) \right] \, ,
\end{equation}
where $\Lambda$ denotes an energy cutoff. 
The decay rate into a pair of charged leptons
$\ell$ is given by Eq.~\eqref{eq:widtt} with $t \to \ell$
and $N_c = 1$.
Therefore, even for an effective ALP--lepton coupling equal to $c_t$, the leptonic partial widths are suppressed by the small lepton masses by at least a factor $7 \times 10^{-5}$,
rendering leptonic decays irrelevant in the
present discussion.


\bibliographystyle{JHEP}
\bibliography{lit}

\end{document}